\documentclass[fleqn,usenatbib]{rasti}

\usepackage{newtxtext,newtxmath}

\usepackage[T1]{fontenc}
\usepackage{ulem}
\DeclareRobustCommand{\VAN}[3]{#2}
\let\VANthebibliography\thebibliography
\def\thebibliography{\DeclareRobustCommand{\VAN}[3]{##3}\VANthebibliography}

\usepackage{graphicx}	
\usepackage{amsmath}	
\usepackage{xcolor}
\usepackage{multirow}

\title[A solid metalon for calibration]{Use of solid fused silica etalon with broadband metallic coatings for calibration of high-resolution optical spectrograph}

\author[S. Ghosh et al.]{
Supriyo Ghosh,$^{1}$ 
\thanks{E-mail: s.ghosh3@herts.ac.uk, supriyoani89@gmail.com}
William Martin,$^{1}$
Kajal Kunverji $^{1}$
and 
Hugh R. A. Jones $^{1}$
\\
\\
$^{1}$ Centre for Astrophysics Research, Department of Physics, Astronomy and Mathematics, University of Hertfordshire, Hatfield, Hertfordshire AL10 9AB, UK \\
}
\date{Accepted XXX. Received YYY; in original form ZZZ}

\pubyear{2025}

\begin{document}
\label{firstpage}
\pagerange{\pageref{firstpage}--\pageref{lastpage}}
\maketitle

\begin{abstract}

Wavelength calibration is a key factor for high-resolution spectroscopic measurements for precision radial velocities. Hollow-cathode lamps (e.g., ThAr), absorption cells (e.g., iodine cell), dielectric coated Fabry-P\'erot etalons and laser frequency combs have been implemented over the years for precise wavelength calibration and wavelength drift measurements. However, due to their various impediments as wavelength calibrators, investigations of alternative methods remain of prime interest. In this paper, we examined the feasibility of low-cost ($\sim$ $\$$1000) commercially available solid fused silica etalon with a broadband metallic coating as a calibrator. We studied the behaviour for two cavity spacings (free spectral range of 1/cm and 0.5/cm) with temperature from theoretical derivation and experimental data. Our setup had a temperature stability of 0.8 mK for a calibrator system using an off-the-shelf dewar flask with active stabilisation. Our result from radial velocity drift measurements demonstrated that such a calibration system is capable of providing higher signal-to-noise calibration and better nightly drift measurement relative to ThAr in the wavelength range between 470 nm and 780 nm. A similar result has been previously found for Fabry-P\'erot etalons, and although the metalon solution lacks the efficiency of an etalon, it does offers a cost-effective broadband solution, which should be less prone to aging relative to complex dielectric mirror coatings. Nonetheless, long-term monitoring is required to understand the metalon behaviour in detail. 
\end{abstract}




\begin{figure*}
	\centering
	\includegraphics[width=6.8in, height=2.5in]{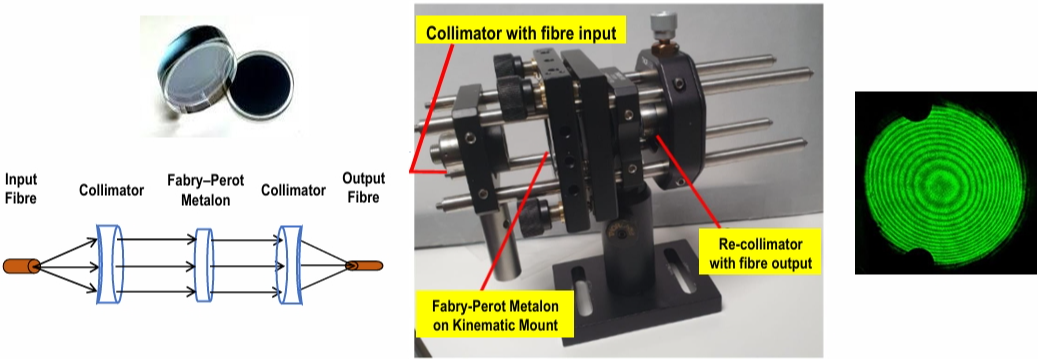}
	\caption{The top left-panel displays OP-8565 solid fused silica metalon from LightMachinery [image origin: \url{https://lightmachinery.com/optics-catalog/etalons-metal-coated/}]. The optical layout of the Fabry-P\'erot metalon from input fibre to output fibre is presented in the bottom left panel. The middle panel shows the assembled cage system of Fabry-P\'erot metalon. Collimator with fibre input, Fabry-P\'erot metalon on kinematic mirror mounts, and re-collimator with fibre output are labelled. The right panel represents the diffraction pattern of the Fabry-P\'erot metalon, illuminated by a green laser of wavelength 532 nm. The image was captured during the optimisation of alignment process with the collimator where a direct laser beam was allowed to pass through the FPM setup. This is not the optimum for the collimator design. In addition, the semi-circle dark cutouts are shadows from the adjuster of the kinematic mounts on the collimator.  
    }  
	\label{Fig:metalon-cagesystem}
\end{figure*}

\section{Introduction}
Over more than a decade, high-resolution spectrograph (HRS) has been devoted to searching for and characterising exoplanets by measuring radial-velocity (RV) shift on the host star spectrum along our line of sight. One of the key factors in this measurement is the accuracy and precision of wavelength calibration (WC) of HRS. WC sets up a physical scale by mapping absolute wavelengths onto the pixel position of the detector. The calibration source plays a key role for any measurement with HRS. The performance of the calibration source directly impacts the accuracy and precision of the measurements. Thus, highly stable and precise calibration sources, having broad wavelength coverage, are fundamental requirements.

Hollow-cathode lamps (HCLs), for example, thorium-argon (ThAr; \citealt{1996A&AS..119..373B, 2000SPIE.4008..582P, 2006SPIE.6269E..0PL, 2007A&A...468.1115L}), thorium-neon (ThNe; \citealt{1979AISAO..11...70B}), uranium-argon (UAr; \citealt{2021JATIS...7c8005S}) emission lamps, or molecular gas absorption cells, for example, iodine cells \citep{1976Natur.260..227B, 1983PhDT.........2K, 1984A&A...134..134K, 2020PASP..132a4503W, 2024A&A...690A.210R, 10.1093/mnras/staf588}, are traditionally used to perform wavelength calibration \citep{1992PASP..104..270M}. A comparative study on the qualitative behaviour of various HCLs used for wavelength calibration in optical and near-infrared was presented in \citet{Sarmiento_refId0}. Achieving a typical accuracy of a few meters per second was reported in literature using both HCLs and gas absorption cells (see; \citealt{2009A&A...507..487M, 2021JATIS...7c8005S, 2024A&A...690A.210R}). However, both the emission and absorption calibration sources suffer non-uniform spacing, large intensity differences and blending of lines. Furthermore, while gas absorption cells, for example, iodine cells have narrow wavelength coverage ($\sim$520-620 nm, \citealt{2024A&A...690A.210R}) with dense absorption lines, HCLs, for example, ThAr have a limited number of reference lines with relatively broad wavelength coverage. In addition, the finite life-time and the chemical impurity (e.g., \citealt{2018SPIE10704E..07N}) make ThAr unsuitable to use as a sole wavelength calibrator in the current-generation of precision RV instruments \citep{2015A&A...581A.117B, 2021AJ....161..252T, 2023AJ....165..156T}. Moreover, ThAr lamps might be discontinued in the market in the near future because of the increased restrictions on the radioactive element thorium and problems with sourcing quality thorium required (Photron Pty Ltd, private communication).


To overcome the limitation of HCLs, new calibrators, for instance, laser frequency combs (LFCs; \citealt{2007MNRAS.380..839M, 2008Sci...321.1335S, 2010MNRAS.405L..16W, 2012Natur.485..611W}) have emerged as a promising alternative for next-generation RV instruments. LFCs provide a comb-like spectrum with equidistant and approximately equal intense narrow lines and thus can be used as a precise frequency reference. A precision of 1$\times$10$^{-18}$ and 1$\times$10$^{-11}$ were achieved in a laboratory environment \citep{doi:10.1126/science.aay3676} and in on-sky performance at an astronomical observatory \citep{Metcalf:19}, respectively, making such calibrator suitable for detection of Earth-size exoplanets (precision requirement $\sim$10 cm s$^{-1}$). However, LFCs are typically expensive, require complex experimental setups, have high maintenance costs and have proven operationally challenging \citep{2024A&A...690A.210R, 10.1093/mnras/stae920} and have so far not delivered on their potential, e.g., \citet{10.1093/mnras/stae920}. 

On the other hand, a Fabry-P\'erot etalon (FPE) can be a simpler and  cost-effective alternative to LFCs. The FPE is basically a stabilized optical resonator, consisting of two parallel and identical partially-reflective mirrors, with a cavity in-between that can produce an interference pattern (a comb-like calibration spectrum, \citealt{1899ApJ.....9...87P}) while illuminated with a collimated beam from a broadband source. The peaks of the comb-like spectrum are the resonant frequencies of the cavity. The range of frequency between successive peaks is the free spectral range (FSR). The width of a peak when the transmittance is 0.5 is the full width at half maximum (FWHM). The ratio between FSR and FWHM is the finesse of a FPE. It measures the sharpness of FPEs peaks. The characteristics of the modulated pattern are tied to physical properties (for example, effective cavity length), wavelength of light and environmental (for example, temperature and pressure) and illumination conditions \citep{2023AJ....165..156T, Jennings:17}. Thus, FPEs can be used in a variety of applications, including spectroscopy, laser tuning, and optical communication systems as a spectral filter. Depending on the kind of application, FPEs can be designed in different ways.

In astronomy, the majority of HRS uses dielectric-coated air-spaced FPE as a calibrator. In such FPEs, reflective mirrors are coated in multiple layers with dielectric having reflectances of $\sim$0.98, and the air-gap is maintained by spacers made from ultra-low-expansion materials. As mentioned above, the spectral properties change with effective cavity length, which in turn depends on the refractive index of cavity material (air for air-spaced FPE). Furthermore, the refractive index is influenced by the environmental conditions of the cavity. Therefore, good stabilisation for such calibrators is required and mostly performed through the passive stabilisation method using a vacuum chamber.

A practical implementation of FPE for precision RVs was made for the HARPS spectrograph \citep{10.1117/12.857951}. Similar calibration systems were developed afterward for various HRS, for example, CARMENES \citep{10.1117/12.926232}, ESPRESSO \citep{2014SPIE.9147E..1HM}, SPIROU \citep{ 2017A&A...601A.102C} and HPF \citep{2021AJ....161..252T}. It is important to note that while the FPE is in use, a secondary calibration source (for example, ThAr) is required to identify the exact wavelength of each peak. Hence, in most observatories, ThAr is used for absolute calibration, and FPE is used for nightly drift measurement. Combining these two calibration sources, a substantial improvement in the accuracy of wavelength calibration was achieved in comparison to using the ThAR lamp alone \citep{2019A&A...624A.122C, 2021AJ....161..258H}. A nightly-drift of 10 cm s$^{-1}$ was reported through optimum stabilisation of the passive air-spaced FPE (\citealt{10.1117/12.857951}, see also \citealt{2014PASP..126..445H, 10.1117/1.JATIS.3.2.025003, 2017A&A...601A.102C, 10.1117/12.2312472, 2019A&A...624A.122C}). Furthermore, \citet{Faria2022} reported an on-sky RV precision of 39 $\pm$ 7 cm s$^{-1}$ using an air-spaced FPE calibrator on the ESPRESSO spectrograph at the Very Large Telescope (VLT).  However, the enhancement of the overall cost and lengthy stabilization period due to the use of vacuum chamber and pumping mechanisms, diminishing the stability of the spectral lines due to the mechanical vibration \citep{PhysRevLett.85.2264}, difficulties in creating very broadband dielectric coatings, wavelength-dependent drift (chromatic behavior) due to the deterioration of dielectric mirror coating \citep{10.1117/12.2312472, 2021AJ....161..252T, 2022A&A...664A.191S, 2025NatAs.tmp...54K} and degradation of the long-term performance due to a very slow drift in etalon temperature have driven interest in exploring new FPE fabrication, for example, solid FPE (\citealt{2023AJ....165..156T}) instead of air-spaced etalons in recent years.
 
In solid FPEs, fused Silica is used as a cavity material, and consequently, bonding between mirrors and spacer is eliminated \citep{2023AJ....165..156T}. They become very compact, relatively cheap, insensitive to air pressure, and more rigid to handle. As they are pressure insensitive due to the solid-cavity, temperature is the main factor stabilising the etalon against the environmental change. \citet{2023AJ....165..156T} demonstrated a precision of $\ge$ 1 ms$^{-1}$ from a thermally controlled solid-etalon. Furthermore, mirrors of the solid etalon are coated with metal on both sides  making them suitable for application over a wide range of wavelengths. In contrast, in dielectric FPE, multiple layers of dielectric materials are used and their optical properties are sensitive to layer density. The metal coating is not highly reflective in thin layers, but they are insensitive to wavelength; hence, very broadband. The main difference between metal and dielectric coated etalons is that metal coating can work over a much larger bandwidth with a linear phase change. The metal coated mirrors are over-coated with a thin layer of optically passive protective dielectric material. The top protective layer may introduce slight modulation, however any such effect is likely negligible compared to the absorption of the metal coating (LightMachinery, private communication). Due to the metal coating, they are familiar as Fabry-P\'erot metalon (FPM). The overall reflectivity will change over time, but this only impacts the finesse. Moreover, such a calibrator is relatively inexpensive ($\sim$ \$1000). In light of all the factors, FPM is believed to be a potential alternative calibrator for HRS, particularly in cost conscious systems. However, its behaviour against the environment and RV performance need to be well-characterized.

In this paper, we present elementary design concepts, feasibility and performance of commercially available solid FPMs operating with active stabilisation. For that, we obtained laboratory data using a high-resolution spectrograph, EXO-planet high-resolution SPECtrograph (EXOhSPEC; \citealt{2019SPIE11117E..0ZL, 2019SPIE11116E..1GK, 2021PASP..133b5001J}) at the University of Hertfordshire (UH) and compared FPM performance with a ThAr lamp. 

The paper is structured as follows: an overview of our FPM system is presented in Section 2. The experimental setup for FPM characterisation is described in Section 3, while Section 4 covers our results and Section 5 presents discussion. Finally, the summary of the study is presented in Section 6.

\section{Overview of our FPM system}
\subsection{Description of FPM}
For our investigation, we have explored commercially available LightMachinery FPM (part number OP-8565-T, T is thickness in mm)\footnote{\url{https://lightmachinery.com/optics-catalog/etalons-metal-coated/}}. These metalons are a silver deposition with SiO$_2$ overcoat for protection against oxidation and moisture and should be robust over time. No degradation has been observed over a year. Nonetheless, the SiO$_2$ protective layer has not been tested on 5-10 year timescales (LightMachinery, private communication). We have used two FPMs of different thicknesses. The first one has a thickness of 3.371 mm (FSR of 1/cm or 30 GHz), hereafter the `thin FPM'. The other has a thickness of 6.743 mm and an FSR of 0.5/cm (15 GHz), the `thick FPM'. Both have a clear aperture of 20 mm. These FPMs have a finesse of about 2.5, transmission $<$ 50\%, and modulation of about 30\% of output signal from 350 nm to 900 nm. The physical properties of FPMs are listed in Table \ref{tab:MetalonDetailed}. The FPM was mounted on a Thorlabs three Adjuster Precision Kinematic Mirror Mount (see, the middle panel of Fig. \ref{Fig:metalon-cagesystem}). 

Two Fibreport collimators (Thorlabs F950FC) were carefully positioned on either side of the FPM to efficiently control the direction of incoming and outgoing light beams as shown in the bottom left-panel and middle-panel of Fig. \ref{Fig:metalon-cagesystem}. The collimators optimise the alignment of the optical components by ensuring that the light rays are parallel to the optical axis. A compact setup needed to be developed to provide a stable and controlled environment for all optical elements along with necessary mounting brackets and holders. The middle panel of Fig. \ref{Fig:metalon-cagesystem} displays the assembled cage system.

\subsection{Alignment}
The precise alignment of all optical components is crucial for the optimal performance of the calibrator. Any misalignment of these components could result in a reduction of calibration accuracy, leading to inaccurate calibration of the spectrograph. Misalignment also results in loss of light intensity that causes signal to noise ratio of the calibration spectrum to deteriorate. To achieve optimal alignment, a green laser of wavelength 532 nm (Thorlabs CPS532) was used. To perform the overall alignment, we removed the FPM from its mount and aligned two collimators. The alignment process was repeated for each collimator in the setup, making sure that each collimator was properly aligned with the laser beam. Thereafter, the FPM was inserted into the setup. Due to the insertion of the FPM, an interference pattern was formed if the setup was aligned. Using the adjusters of the FPM mount, the quality of the interference pattern produced by the whole setup can be improved. Careful adjustments were then made to optimise the visibility of the diffraction pattern (see, the right-panel of Fig. \ref{Fig:metalon-cagesystem}), ensuring that each component was aligned to achieve the desired performance of the calibration device.

A laser power meter was also used to measure the total intensity of light emitted from the output fibre. The positioning of the FPM was adjusted to align the light from the fibre with the axis of the FPM and maximize the amount of light passing through the spectrograph for improved image quality.

The alignment of the optical components underwent rigorous testing until achieving a satisfactory level of image quality, i.e. a modulation of about 30$\%$ interference pattern with precise centering.

\begin{table}
	 \centering
	\caption{Specifications of solid FPMs.}
	\label{tab:MetalonDetailed}
         \resizebox{0.46\textwidth}{!}{%
	   \begin{tabular}{lcccll}
          \hline
		Parameters & Thin & Thick \\
                       & FPM  & FPM  \\
		\hline
           Material & Fused silica & Fused silica \\
           Thickness of FPM ($Z_0$) & 3371 $\mu$m & 6743 $\mu$m \\
           Finesse (\textbf{\textit{F}}) & $~$ 2.5 & $~$ 2.5 \\
           Free Spectral Range (FSR) & 1/cm & 0.5/cm   \\
           Wavelength range & 350 -- 900 nm & 350 -- 900 nm\\
           Reflectivity of each surface of the etalon (R) & $<$ 50$\%$ & $<$ 50$\%$  \\
           $\frac{\partial{Z}}{\partial{T}}$ & 5.10 $\times$ 10$^{-7}$ /$^o$C & 5.10 $\times$ 10$^{-7}$ /$^o$C \\
           $\frac{\partial{n}}{\partial{T}}$ & 8.62 $\times$ 10$^{-6}$ /$^o$C & 8.62 $\times$ 10$^{-6}$ /$^o$C \\
		\hline
	   \end{tabular}}
\end{table}

\subsection{FPM Transmission}
\begin{figure}
	\centering
	\includegraphics[width=3.4in, height=2.5in]{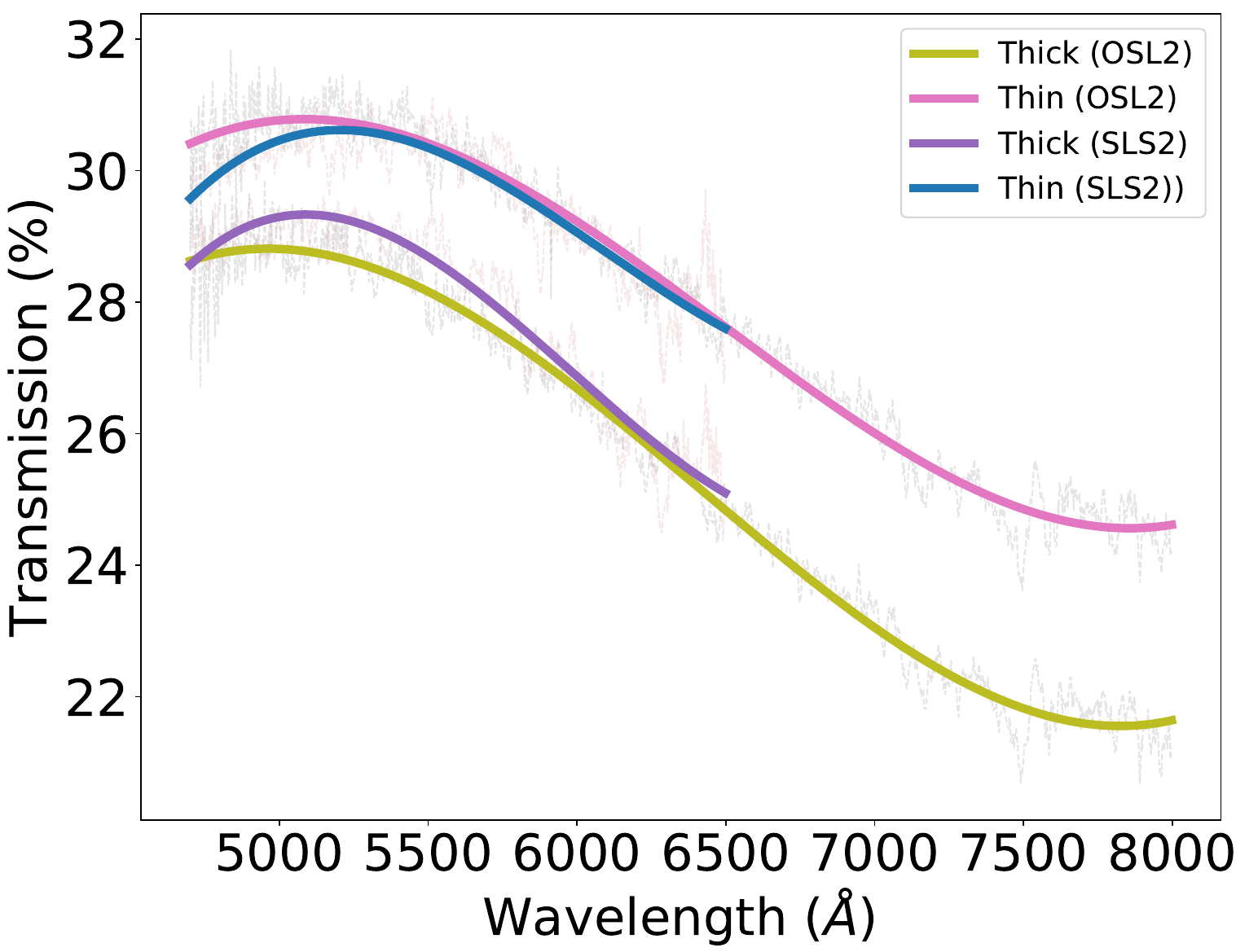}
	\caption{Transmission for both thick and thin FPMs for two light sources, OSL2 and SLS201L/M (abbreviated as SLS2). The faint grey and brown dashed lines represent the observed data points for OSL2 and SLS2 lamps, respectively. Smoothed versions of the transmission curves obtained by applying a Savitzky-Golay filter, are overlaid in solid lines. The truncation of SLS2 at 6500 \AA~is due to low lamp emission at longer wavelengths.}  
	\label{Fig:FPM_Transmission}
\end{figure}

The average transmission of the FPM system was measured by the Ocean optics HR4 high-resolution spectrometer. We used two white lamps, Thorlabs OSL2 and SLS201L/M for our measurements. A combination of two 50 $\mu$m core fibres was used in this measurement, serving both as the input and output of the FPM cage system. In each case, we first recorded the lamp spectrum alone, which serves as a reference spectrum. Then, the FPM was placed into the cage system and the same spectrum was transmitted through the FPM. The ratio of these two spectra measures the average transmission of the FPM. The measured average transmittances for both thick and thin FPM systems are presented in Fig. \ref{Fig:FPM_Transmission} as a function of wavelength for both lamps. Both our FPM and OSL2 are broadband. In addition, OSL2 offers the control of output illumination intensity. Therefore, we used the OSL2 lamp for our further investigation. At 6650 \AA, the transmittance values are about 27\% (24\%)  for thin (thick) FPMs, respectively. As mentioned in \citet{10.1117/12.857951}, this value has to be multiplied by the finesse of the FPM to get the peak-transmittance. 
The quoted finesse by the manufacturer is 2.5 for both thin and thick FPMs as presented in Table \ref{tab:MetalonDetailed}. Therefore, the average transmissions of the FPM are about 67\% (60\%) in that wavelength for thin (thick) FPMs, respectively, which are consistent with quoted values.

\subsection{FPM Environment control}
The cavity spacing for a solid metalon is believed to be independent of wavelength and should mean the cavity spacing should remain constant at any moment in contrast to the dielectric etalon, where the optical thickness of the coating depends on the wavelength of light. This is because the rate of change of refractive index for silica at  wavelengths of our interest is too small at a particular temperature and the interference mostly happens between the metal surfaces, ignoring a very small effect by the outer protective layer. In addition, the skin depth of metal coatings varies with wavelength. For silver, it ranges from $\sim$12 nm at 350 nm to $\sim$20 nm at 900 nm \citep{articleLee2007}. However, this difference is swamped by other influences on FSR such as the dispersion change from the substrate material (LightMachinery, private communication). Nevertheless, in a solid fused silica metalon, the cavity spacing changes with temperature due to the thermal expansion of silica. On the other hand, the refractive index of silica is strongly temperature and wavelength dependent in the optical regime (see, \citealt{Malitson:65}). The wavelength dependence can be derived from the Sellmeier equation. Hence, the important parameter that needs to be derived is how the FPM system drifts with temperature and the required stability of the FPM system that needs to be achieved for 1 m/s RV precision. Nevertheless, a stable and controlled environment is required around the FPM to achieve the optimised performance. As mentioned above, a solid FPM is insensitive to atmospheric pressure; the only environmental parameter to be controlled is the temperature. A thermally insulated enclosure was built to house the overall FPM cage system. We then applied an active stabilization method using a Thermoelectric Cooler (TEC) controller through a tuned integrated proportional-integral-derivative (PID) loop. We investigated two designs of environmental enclosures that used the same temperature control and stabilisation system, Metalon v1.0 and v2.0.

\subsubsection{Metalon v1.0}
Metalon v1.0 was a 3D printed thermally insulated enclosure. This enclosure has two chambers inside to provide sufficient insulation. A cut-out hole of diameter about 2 mm was made to allow fibres entrance and exit. As an insulating material, acrylic was preferred due to its high thermal resistance and durability compared to glass. A layer of gold foil was carefully applied on the inner side of the inner and outer chambers to improve shielding against infrared radiation \citep{Garoli:15}. To enhance insulation further, space blankets were added at the outside of the two chambers. Aerogel and black foam were placed at the gap in between two chambers. Another piece of aerogel was placed at the bottom of the inner chamber for the cage system to sit on as well as to protect the inner gold leaf from any sort of scratch.

\begin{figure}
	\centering
	\includegraphics[width=3.2in, height=3.5in]{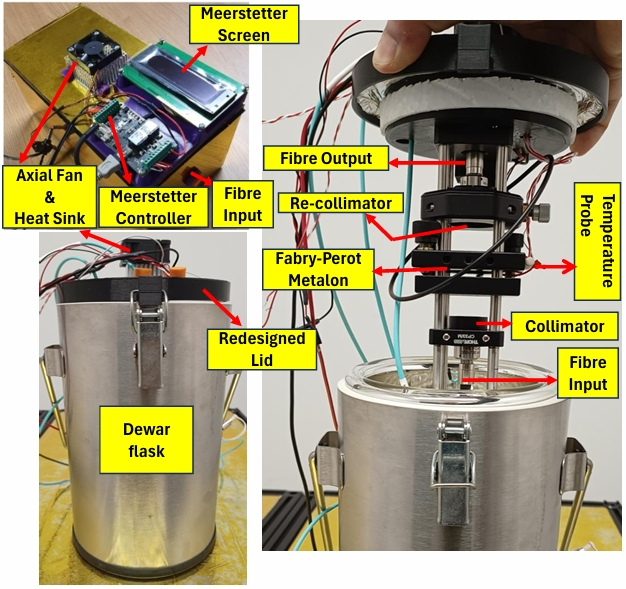}
	\caption{The top left panel displays assembled Acrylic enclosure for Metalon v1.0. The bottom left panel shows a dewar flask with a redesigned lid for Metalon v2.0 insulation. The right panel presents FPM cage system with the dewar flask.}  
	\label{Fig:metalon-enclosure1}
\end{figure}

The lid of the enclosure was designed to include several components to control temperature. These components included a small internal fan and heat sink with air circulation, a relatively large external fan for increased airflow over the external heat sink, a Peltier module, a sensor for temperature measurement, a Meerstetter controller for regulating temperature, and a Meerstetter screen to display temperature readings. The catalog number of each component is listed in Table \ref{tab:TEC_details}. The lid is also equipped with a hole of about 6 mm to allow wires to pass. This includes the temperature sensor and small fan, which are routed inside the box. We have directly measured vibrations from the magnetic bearing fans with an interferometer in the enclosure and found that the vibration is too small to have any impact on the FPM operation. The development of such an enclosure is illustrated in Fig. \ref{Fig:metalon-enclosure1}. We achieved a stability of 4 mK (discussed in detail in sec \ref{sec: MetalonRVPerformance}) using this 3D printed version with active stabilization (presented in sec \ref{sec:active_stabilisation}).

\begin{table}
	 \centering
	\caption{Components of TEC with their corresponding catalog numbers.}
	\label{tab:TEC_details}
         \resizebox{0.46\textwidth}{!}{%
	   \begin{tabular}{lcc}
          \hline
		Component & Catalog number & Characteristic \\
           \hline
           DC Axial Fan (Small) & Farnell - `3886330' & 5 V \\
           DC Axial Fan (Large) & Farnell - `1217970' & 12 V \\
           Large Heat Sink & RS - `674-4838' & 2.4K/W \\
           Peltier Module & RS - `2490-1339' & 21.2W, 3.9A, 8.8V \\
           Small Heat Sink & RS - `184-6718' & 24.5K/W \\
           Thermistor & Farnell - `725-5743' & \\
		\hline
	   \end{tabular}}
\end{table}

\subsubsection{Metalon v2.0}
It was evident from our preliminary investigation that an improved enclosure was needed to achieve 1 m/s RV stability (see Sec \ref{sec: TemperatureSensitivityMeasurements}). For that, we looked for an off-the-shelf solution, and a readily available Dilvac dewar flask (MS222) was applied. This stainless steel container is designed for safe use and transport of liquid gases. The lid of the container was redesigned and 3D printed to house all aforementioned components for active stabilisation. Fig. \ref{Fig:metalon-enclosure1} shows the Dewar system with a modified lid. This housing system with the same temperature controller achieved a stability of 0.8 mK, as discussed in sec \ref{sec:active_stabilisation}. This version of the FPM system is called Metalon v2.0. 

\subsubsection{TEC controller and PID calibration} \label{sec:active_stabilisation}
To enhance the performance of the thermal enclosure and to control the temperature precisely, a controlled loop feedback system was deployed by the TEC. TEC works based on the performance of a PID controller. A PID controller consists of three components: proportional (P) control, integral (I) control, and derivative (D) control. It starts the operation by comparing the desired setpoint temperature with the actual value, which is measured by a sensor placed adjacent to the FPM inside the thermal enclosure and an error signal is being generated. The overall signal generated by these three controllers was then sent to the actuator to bring the error signal down. For optimal performance, the parameters that control PID loop, however, need to be optimised by tuning the system. We first auto-tuned the PID controller using the TEC software to get the initial values. From there, we tuned it further manually by trial and error adjusting the values of three control gains. We continued this process until we obtained the desired overall response from the PID controller.

\subsection{Fabry-Perot Metalon Theoretical Approach}
A simple analysis of the maxima in transmission of a (solid) etalon with refractive index $n$, distance between the mirrors of $Z$, and mirror reflectivity $R$ is developed. 
The phase difference between successive reflected beams is
		$\delta = (2\pi/\lambda) * 2nZ$.  
The maxima occur when $m\lambda = 2nZCos(\theta)$, where m is an integer and $\theta$ is the angle of incidence. If $\theta = 0$ i.e. for normal incidence $m\lambda = 2nZ$.

The wavelength difference between two adjacent maxima or minima can be found by the following:
\begin{align}
m\lambda^\prime & = m\lambda + \lambda \nonumber\\
m(\lambda^\prime - \lambda) & = \lambda \nonumber\\ 
\Delta\lambda & = \lambda/m \nonumber
\end{align}
where  $\lambda^\prime - \lambda = \Delta\lambda$ is termed as Free Spectral Range ($FSR$). Thus,
\begin{equation}\label{eq:FSR_equation}
FSR = \Delta\lambda = \lambda/m = \lambda^2/(2nZ)
\end{equation}	

\begin{figure*}
	\centering
	\includegraphics[width=3.5in, height=2.6in]{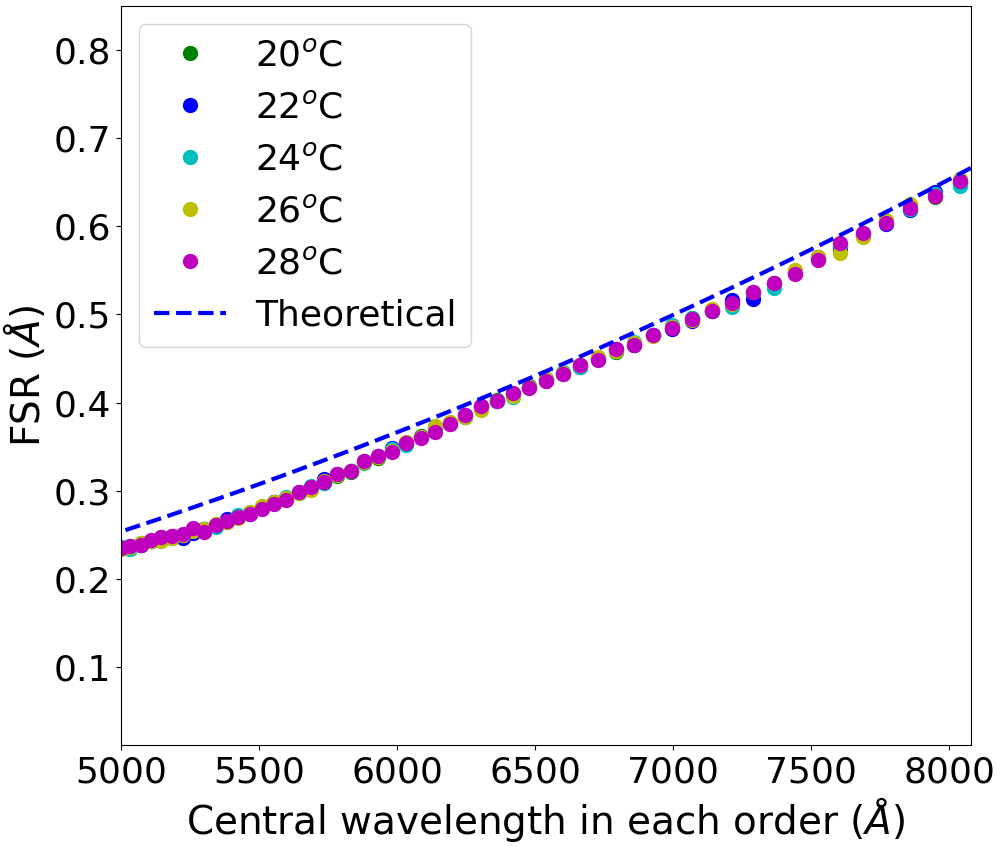}
	\includegraphics[width=3.5in, height=2.6in]{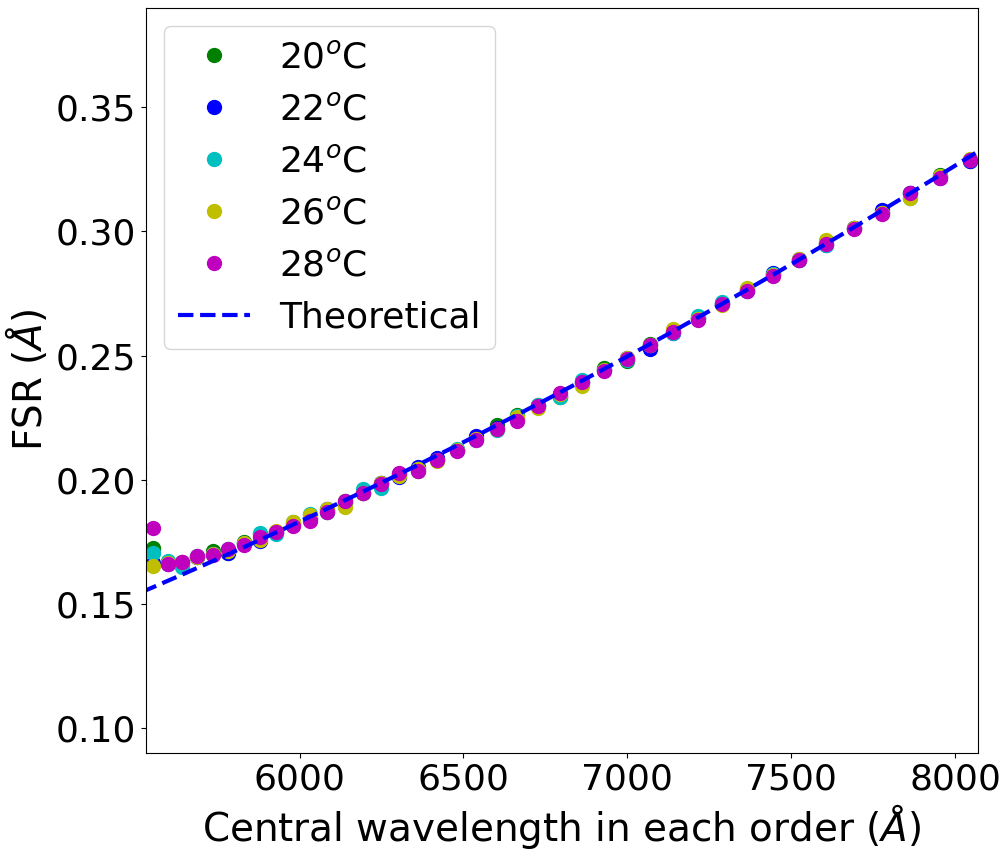}
	\caption{FSR comparison between theoretical and measured values for both thin (left) and thick FPMs (right). The experimental FSR was evaluated for five different temperatures. For thin FPM, the theoretical curve does not follow the experimental measurements. However, for the thick FPM, our theoretical and experimental measurements match well.}
     \label{Fig:FSR_comparison}
\end{figure*}

The measured theoretical FSR is presented later in this paper (see Fig. \ref{Fig:FSR_comparison}). Looking at temperature variation, assuming both n and z have temperature coefficients, from equation \ref{eq:FSR_equation}, we get
\begin{align}  \label{eq:fsr_calc}
\dfrac{d}{dT}\;(FSR) & = \frac{\lambda^2}{2nZ}(-\frac{1}{Z}\frac{\partial{Z}}{\partial{T}} - \frac{1}{n}\frac{\partial{n}}{\partial{T}}) \nonumber\\ 
&= FSR*(-\frac{1}{Z}\frac{\partial{Z}}{\partial{T}} - \frac{1}{n}\frac{\partial{n}}{\partial{T}}) \nonumber\\ 
\Delta FSR &= FSR*(-\frac{1}{Z}\frac{\partial{Z}}{\partial{T}} - \frac{1}{n}\frac{\partial{n}}{\partial{T}})*\Delta T \nonumber\\ 
FSR^\prime &= FSR*[1+(-\frac{1}{Z}\frac{\partial{Z}}{\partial{T}} - \frac{1}{n}\frac{\partial{n}}{\partial{T}})*\Delta T] 
\end{align} 
where  $\Delta FSR = FSR^\prime - FSR$ is change in FSR for a temperature of $\Delta T$. Note that as the temperature increases, the FSR decreases. 

The wavelength shift at a particular wavelength (order) requires an additional calculation. Note that each order is shifted by $\Delta$FSR, but the total shift at a particular wavelength will depend on the actual FPM order ($m$). Assuming that we are working with the same order when measuring the shift, we have
\begin{align} \label{eq:shift_derivation1}
m\lambda &= 2nZ \nonumber\\
m\lambda^\prime & = 2(nZ +\Delta(nZ))\nonumber \\ 
\frac{\lambda^\prime}{\lambda} & = \frac{nZ + \Delta (nZ)}{nZ} \nonumber \nonumber \\
\lambda^\prime & = \lambda*\frac{nZ + \Delta (nZ)}{nZ} \nonumber \\
\lambda^\prime - \lambda & = \lambda*\frac{\Delta (nZ)}{nZ}
\end{align}
Ignoring the rate of change of refractive index with $\lambda$ term (i.e. $\frac{\partial{n}}{\partial{\lambda}}$), we get
\begin{align}\label{eq:shift_derivation2}
d(nZ) = (n*\frac{\partial{Z}}{\partial{T}} + Z*\frac{\partial{n}}{\partial{T}}) dT \nonumber\\
\Delta (nZ) = (n*\frac{\partial{Z}}{\partial{T}} + Z*\frac{\partial{n}}{\partial{T}}) \Delta T
\end{align} 

From equations (\ref{eq:shift_derivation1}) and (\ref{eq:shift_derivation2}), we get
\begin{align}\label{eq:shift_derivation_final}
\delta \lambda & = \lambda^\prime - \lambda \nonumber \\
     &= \lambda*(\frac{1}{Z}*\frac{\partial{Z}}{\partial{T}} + \frac{1}{n}*\frac{\partial{n}}{\partial{T}}) \Delta T
\end{align}  

\begin{figure*}
	\centering
	\includegraphics[width=3.5in, height=2.55in]{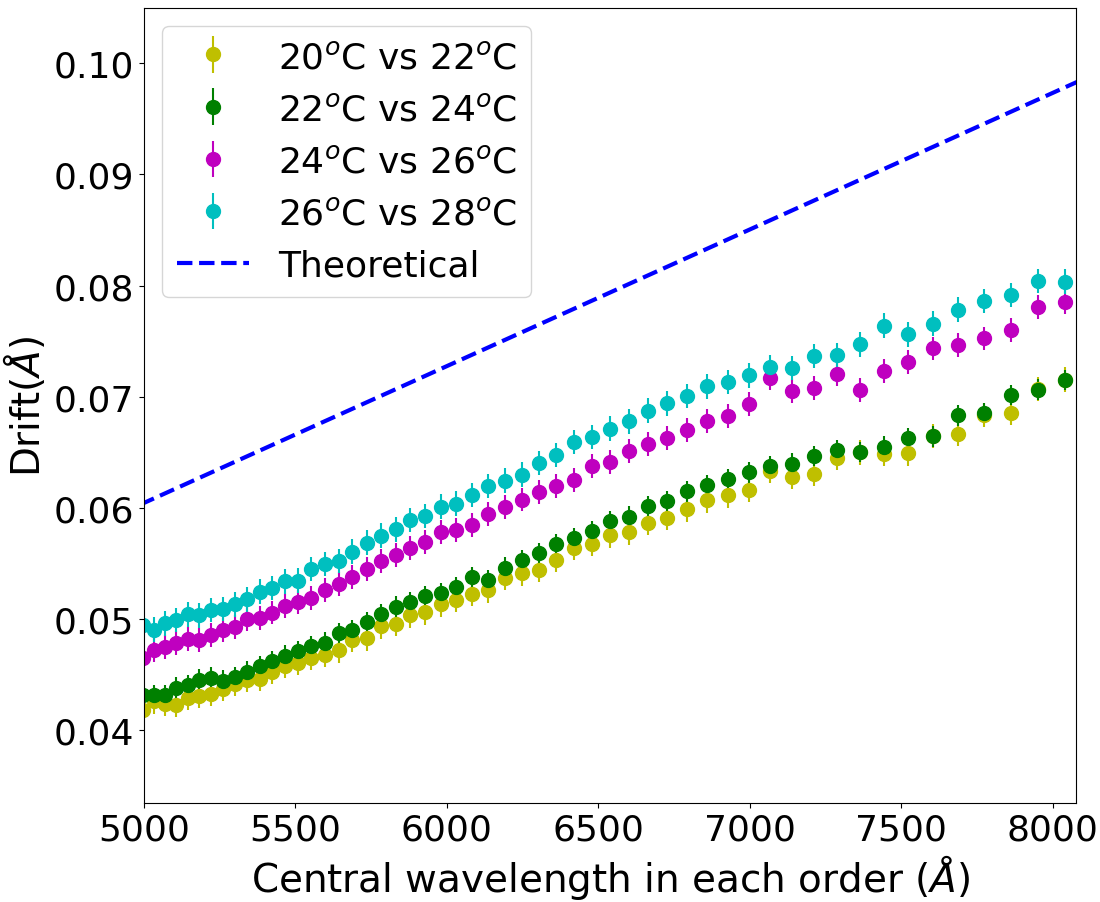}
	\includegraphics[width=3.5in, height=2.55in]{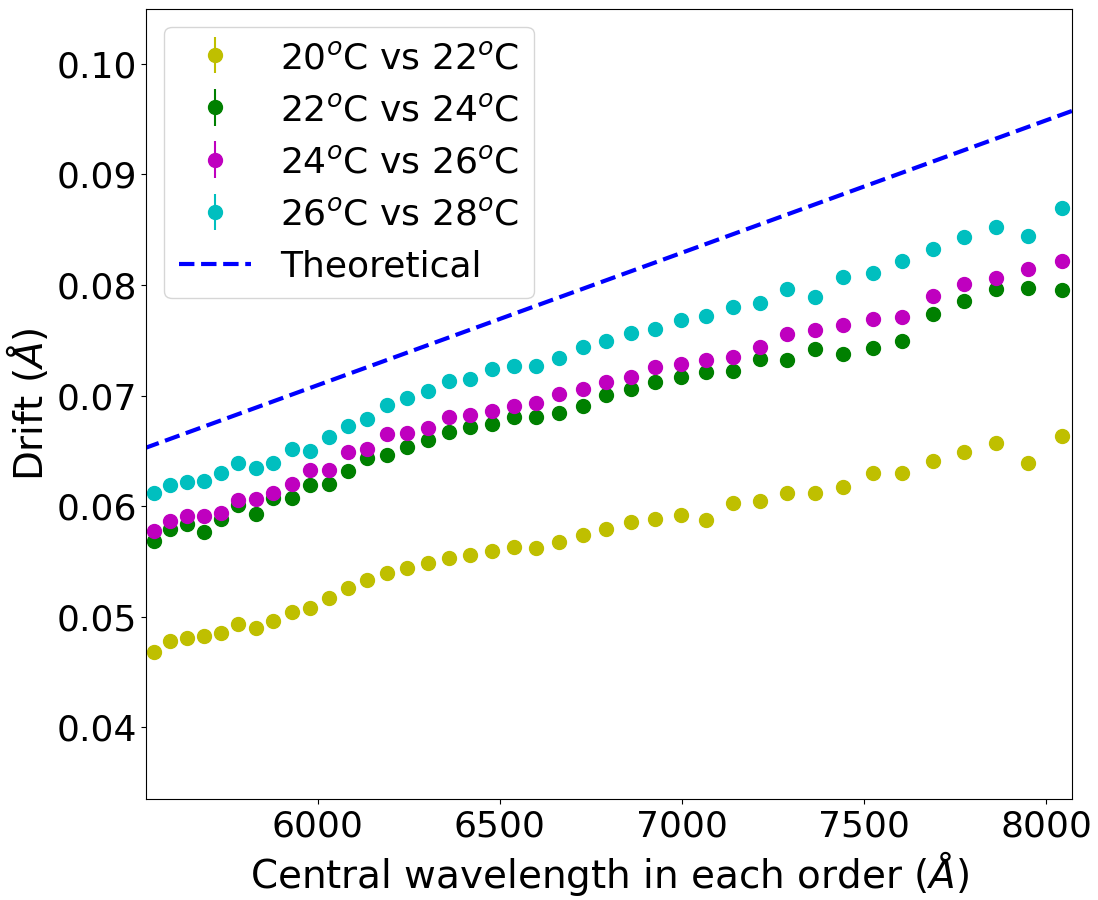}
	\caption{Temperature drift comparison between theoretical and measured values for both thin  (left) and thick (right) FPMs. The experimental drift was evaluated for four different cases from 20$^{\circ}$C to 28$^{\circ}$C in steps of 2$^{\circ}$C. }\label{Fig:drift_vs_wavelength_for_temperature_change}
\end{figure*}

Also, the Sellmeier equation for fused silica is given by
\begin{equation}
n^2 = 1 + \sum_{i=1}^{3} \frac{B_i\lambda^2}{\lambda^2 - C_i}
\end{equation}
where, $\lambda$ is in $\mu$m and B$_1$ = 0.6961663 $\mu$m$^{-2}$, B$_2$ = 0.4079426 $\mu$m$^{-2}$, B$_3$ = 0.8974794 $\mu$m$^{-2}$, C$_1$ = 0.00467914826 $\mu$m$^2$, C$_2$ = 0.0135120631 $\mu$m$^2$, and C$_3$ = 97.934002538 $\mu$m$^2$ \citep{Malitson:65}. The equation provides wavelength dependent refractive index of the fused silica for our theoretical derivation. The measured theoretical drift based on eq. \ref{eq:shift_derivation_final} is presented later in the paper (see, Fig. \ref{Fig:drift_vs_wavelength_for_temperature_change}). It is noted that the absolute value of refractive index at a given wavelength varies with temperature as well as $\frac{\partial{n}}{\partial{T}}$ changes as a function of wavelength, which we ignored in our above drift measurements. However, if we evaluated drifts considering wavelength and temperature dependent values of $n$, $\frac{\partial{n}}{\partial{\lambda}}$ and $\frac{\partial{n}}{\partial{T}}$ for 2$^{\circ}$C temperature change at 22$^{\circ}$C and 27$^{\circ}$C from \citet{leviton2008temperaturedependentabsoluterefractiveindex}, we found that the difference in drift would be about 10$^{-6}$ \AA~ at 6500 \AA, suggesting negligible impact on our above theoretical calculation.

\begin{figure*}
	\centering
	\includegraphics[width=6.2in, height=2.5in]{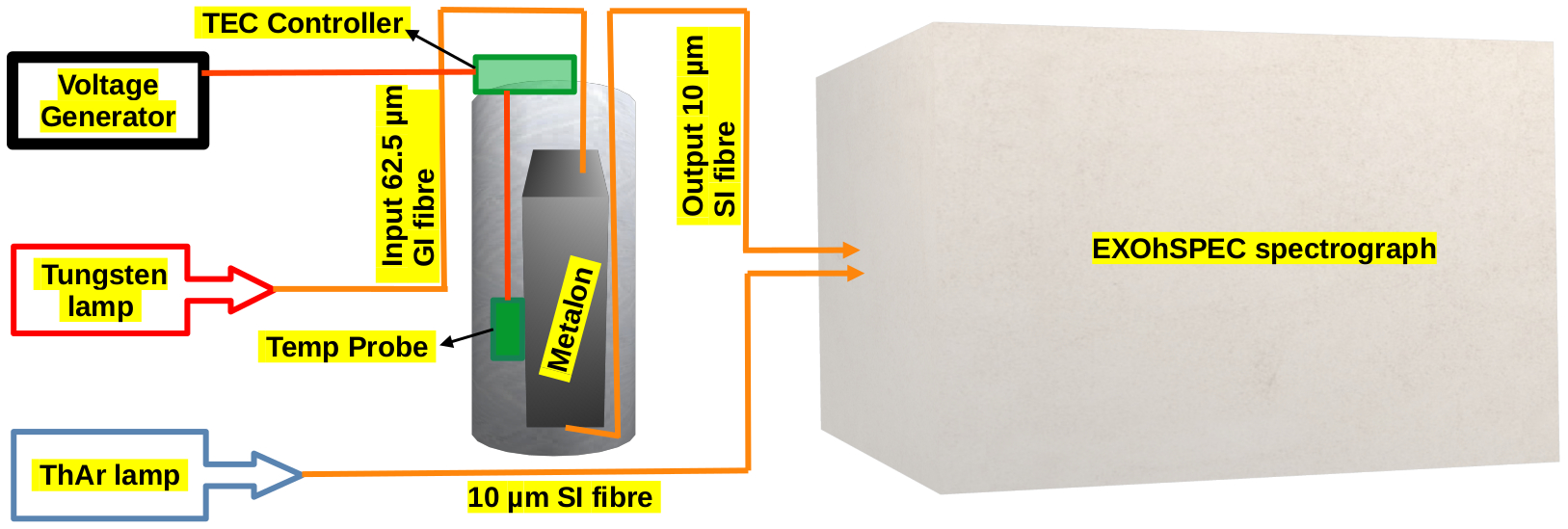}
	\caption{Schematic layout of the overall experimental set-up to envisage the performance of a metalon. The overall setup was similar for Metalon v1.0 and Metalon v2.0. The only difference was that we used our in-house developed thermal enclosure for Metalon v1.0 and dewar flask for Metalon v2.0 to house the FPM system.}
     \label{Fig:metalon-experimental-setup}
\end{figure*}
\begin{figure*}
	\centering
	\includegraphics[width=6.5in, height=2.5in]{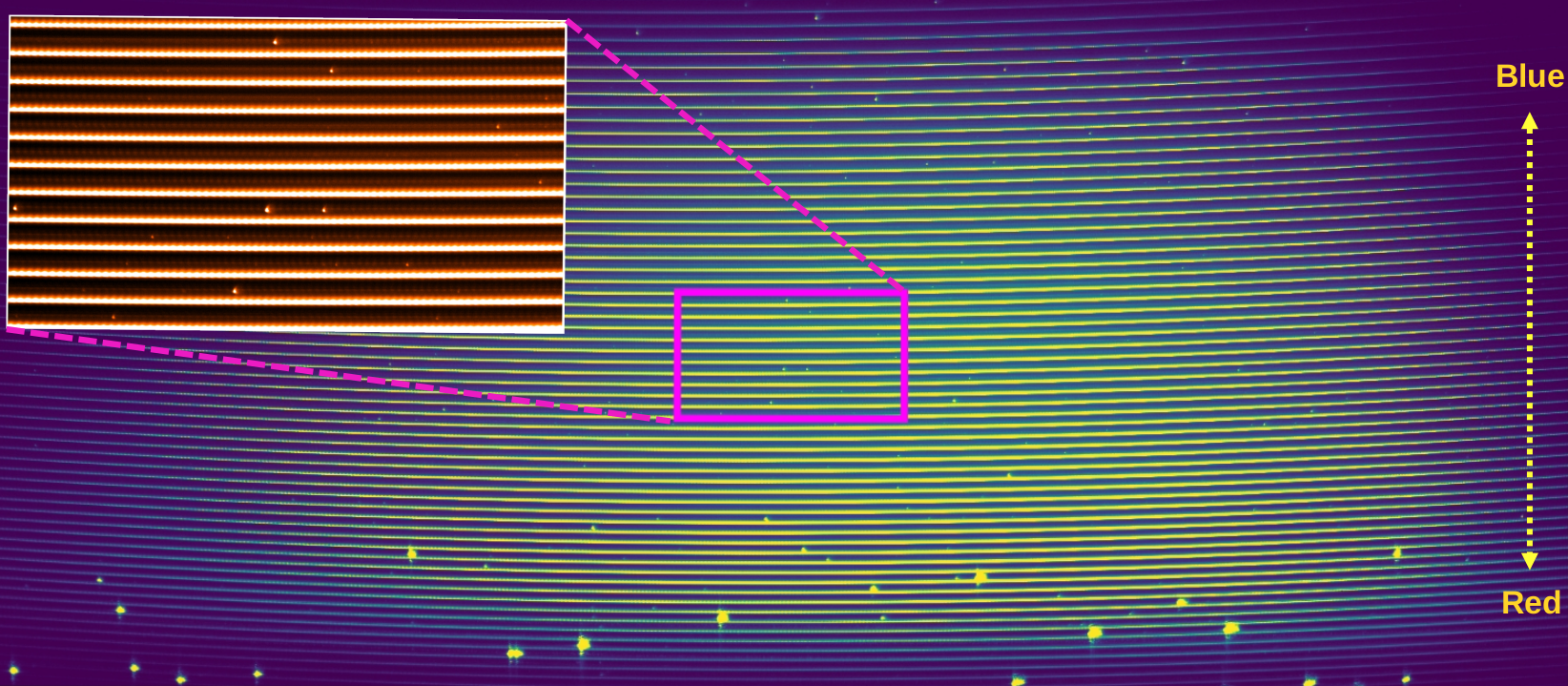}
	\caption{A raw frame displaying FPM spectrum and ThAr lamp spectrum side by side obtained using EXOhSPEC for an exposure of 100 s.}
     \label{Fig:metalon-spectrum_comparison}
\end{figure*}
\section{Experimental set-up and data reduction} \label{Sec:Experimental_setup_and_Spectrum_extraction}
 We examined the behaviour as well as the performance of FPMs on the UH prototype of the EXOhSPEC spectrograph. The EXOhSPEC is a fibre-fed spectrograph designed for radial velocity studies of exoplanets providing a wavelength range of 460 -- 880 nm at a spectral resolution of about 82500  (Table 2, \citealt{2021PASP..133b5001J}). A precision of 3.5 milli-pixels ($\sim$ 4 m/s) was achieved in a standard laboratory environment \citep{2021PASP..133b5001J} using an active stablisation system. For some of the experiments reported here, we control the temperature of the spectrograph but did not use the active stablisation system. We used both thick and thin FPMs for our experiment. The specifications of these FPMs are listed in Table \ref{tab:MetalonDetailed}.
 
 The input of the FPM was connected to a white light source (broadband halogen lamp, Thorlabs OSL2) through a graded-index fibre of 62.5 $\mu$m core. A bifurcated fibre was used to feed the spectrograph. For simultaneous FPM and ThAr observations, one breakout leg of the bifurcated fibre was connected to the output end of the FPM through a step-index fibre of 10 $\mu$m core -- the so-called science fibre. The other leg was connected to the ThAr lamp with a similar fibre -- the so-called calibration fibre. At the beginning of the experiment, the science fibre was fed by tungsten and ThAr lamps one after another to trace the cross-dispersed orders and determine the wavelength solution, respectively, while the calibration fibre was in dark mode. A similar acquisition of frames was followed for calibration fibre while the science fibre was in dark mode. Then, the simultaneous FPM and ThAr observations were carried out by feeding the science and calibration fibres with the FPM and ThAr, respectively.
 
 The thermal enclosure of FPM was kept outside of the spectrograph enclosure, and two sensors (Labfacility 010010TD) were used to measure the temperature of the internal air of the spectrograph and the FPM enclosures. For initial investigation, we explored Metalon v1.0 and found that we need to improve the enclosure of the FPM system to achieve a RV stability of 1 m/s (see Sec \ref{sec: TemperatureSensitivityMeasurements}). Moving from 3D printed version to dewar flask, we carried out experiments only to test the RV performance of Metalon v2.0. Our experimental setup for Metalon v2.0 is illustrated in Fig. \ref{Fig:metalon-experimental-setup}. The setup for Metalon v1.0 was exactly similar to Metalon v2.0 except the thermal enclosure housing the FPM system.

To measure the temperature sensitivity of both thin and thick FPMs (Metalon v1.0), we illuminated the FPMs with a broadband source (tungsten lamp - Thorlabs SLS201L/M) and varied set-point of the FPM box temperature from 20$^{\circ}$C to 28$^{\circ}$C in steps of 2$^{\circ}$C. We obtained 5 frames, each with 100 s exposure, for any set temperature using the EXOhSPEC spectrograph. A raw frame of conjugate FPM and ThAr spectra is shown in Fig. \ref{Fig:metalon-spectrum_comparison} and the relative richness of FPM pattern can be seen in comparison with ThAr.

\begin{figure}
	\centering
	\includegraphics[width=3.3in, height=2.7in]{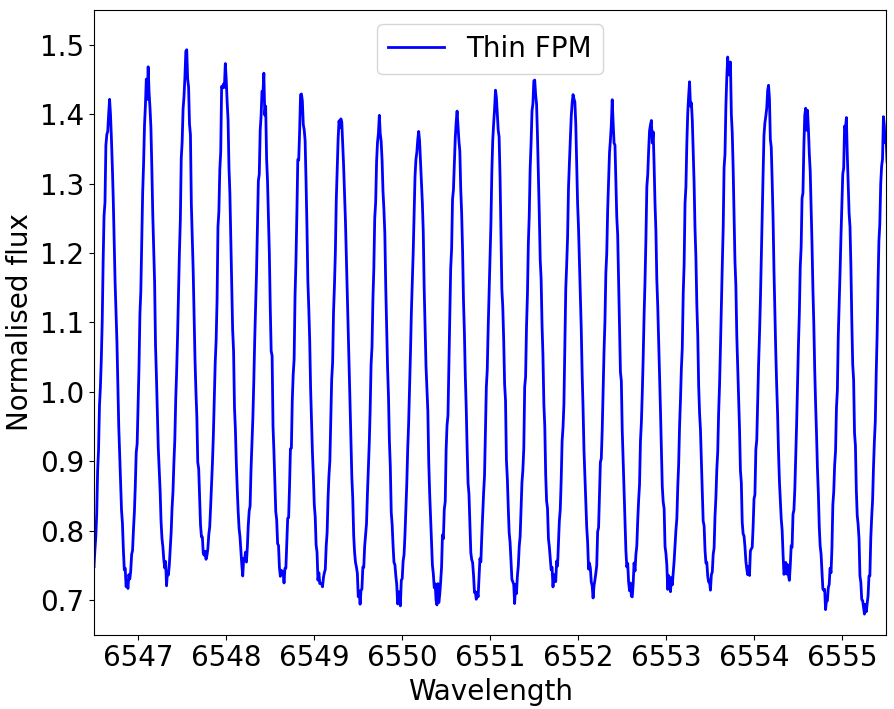}
         \includegraphics[width=3.3in, height=2.7in]{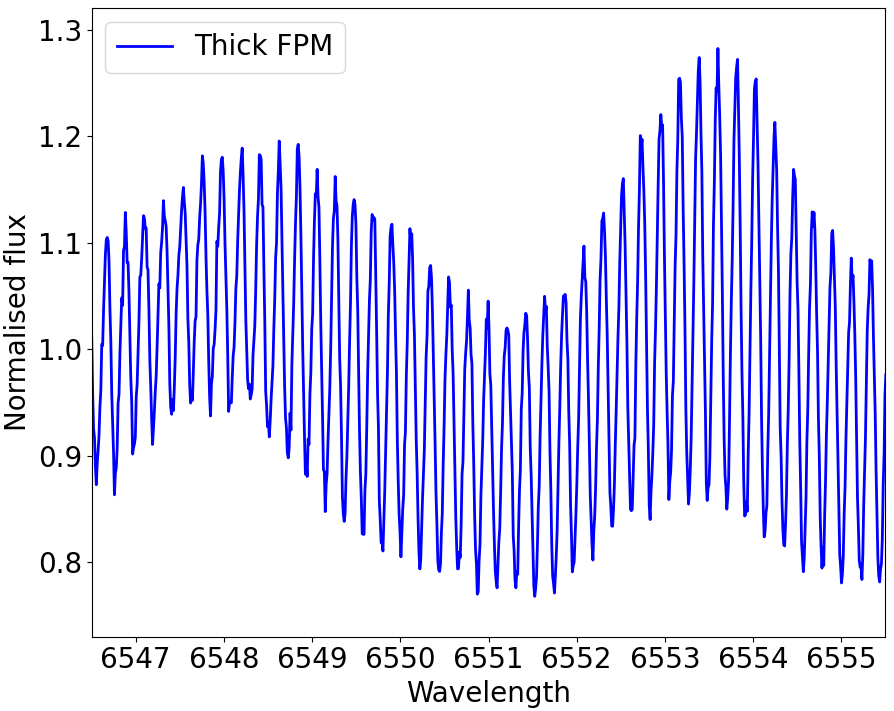} 
	\caption{A portion of extracted thin (FSR: 1/cm) and thick (FSR: 0.5/cm) FPMs spectra around 6550 \AA~ is displayed. It can be noted that the observed peaks are twice denser for the thick FPM than the thin.}
	\label{Fig:extracted_spectrum}
\end{figure}

All data were analyzed in the Python-based data reduction pipeline HiFLEx \citep{2020PASP..132f4504E}. The raw dataset is first corrected for bias, dark, and flat. Following the standard data reduction procedure, 1D spectra of ThAr and FPM were extracted. ThAr lines were used for absolute wavelength calibration. The wavelength calibrations of FPM and ThAr spectra based on simultaneous observations were performed using the individual wavelength solutions of the science and calibration fibres, respectively, determined at the beginning of the experiment. A portion of extracted spectra for both thin and thick FPMs is shown in Fig. \ref{Fig:extracted_spectrum}. The result is discussed in Secs \ref{sec: FreeSpectralRangeMeasurements} and \ref{sec: TemperatureSensitivityMeasurements}. Fig. \ref{Fig:extracted_spectrum} demonstrated that the signal-to-noise ratio (SNR) was much better for the thin FPM than the thick FPM for an exposure of 100 s. Therefore, for practical reasons, we only carried out our test for RV performance with the thin FPM.

To investigate the RV performance of the FPM, we carried out simultaneous FPM and ThAr observations as mentioned above at a temperature that is close to the temperature of a typical room. Thus, the temperatures of the FPM box as well as the spectrograph were set at 24$^{\circ}$C and 23$^{\circ}$C, respectively, for the Metalon v1.0. We continued our experiment for approximately 24 hours. In this time interval, we obtained multiple frames for 7 sets of exposure runs. The exposure for each frame was 100 s. We took 40 frames in each set for the ﬁrst four sets of exposures. The exposure of the first set was started immediately after setting the FPM box temperature up. The time interval between the last frame of one set and the first frame of the following set is about 1 hour. The 5$^{th}$ set consists of 120 frames with an interval of 30 s between two consecutive frames, which essentially gives continuous monitoring of FPM behaviour for about 4 hours. The experimental set-up was left in operation overnight, and the 6$^{th}$ set was taken in the following morning. For this case, 65 frames (30 s intervals between each exposure) were obtained. In the 7$^{th}$ set, we obtained 10 frames with the same exposure time in each frame. This  experiment helped to investigate both short timescale ($\sim$ 1 hour for a single set of 40 exposures) as well as relatively longer timescale variations. The result is discussed in Sec \ref{sec: MetalonRVPerformance}. At the same time, we recorded the temperature variations of the spectrograph and FPM box.

To analyze the RV performance of the Metalon v2.0, we carried out our experiment continuously for 12 hours. Over this period, we obtained multiple frames with an exposure of 100 s. Furthermore, the experiment with Metalon v2.0 was carried out in a general purpose laboratory with the air conditioner off, while the experiment with Metalon v1.0 was performed with the air conditioner on. The air conditioning of the laboratory was extensively controlled by the centralized air conditioning system of the whole building. Moreover, we have noticed an air-con cycle of a period of 30 mins from monitoring temperature in our laboratory, which might impact our result. Therefore, we kept the air conditioner off while experimenting with Metalon v2.0. 

To derive the frame-to-frame RV drift for both FPM and ThAr spectra for both Metalon v1.0 and v2.0, we used standard RV software, Template-Enhanced Radial velocity Re-analysis Application (TERRA; \citealt{Anglada-Escude_2012}) and SpEctrum Radial Velocity AnaLyser (SERVAL, \citealt{refId0}) through our data reduction pipeline HiFLEx after the extraction and calibration of the spectrum (see, \citealt{2020PASP..132f4504E}). Both TERRA and SERVAL measured RVs using the least-squares minimization technique between each observed spectrum and a high SNR template created from the same observations. While SERVAL used all available observations to create the template spectrum, TERRA considered only the best 100 observations in regards to the SNR to build the template. As the barycentric correction is intrinsic to the HiFLEx pipeline before RV analysis, we need to perform this correction although the FPM setup and the spectrograph are relatively stationary. We treated both FPM and ThAr as the Sun to make the pipeline work without significant modification. The pipeline used barycorrpy \citep{Kanodia_2018} to perform barycentric correction. The RVs of two consecutive FPM spectra were subtracted to measure the temporal RV drift. A similar method was followed to measure the temporal RV drift for ThAr. Therefore, all simultaneous FPM and ThAr observations provide the relative comparison of temporal RV drifts between FPM and ThAr.

\begin{figure*}
	\centering
	\includegraphics[width=2.2in, height=2.2in]{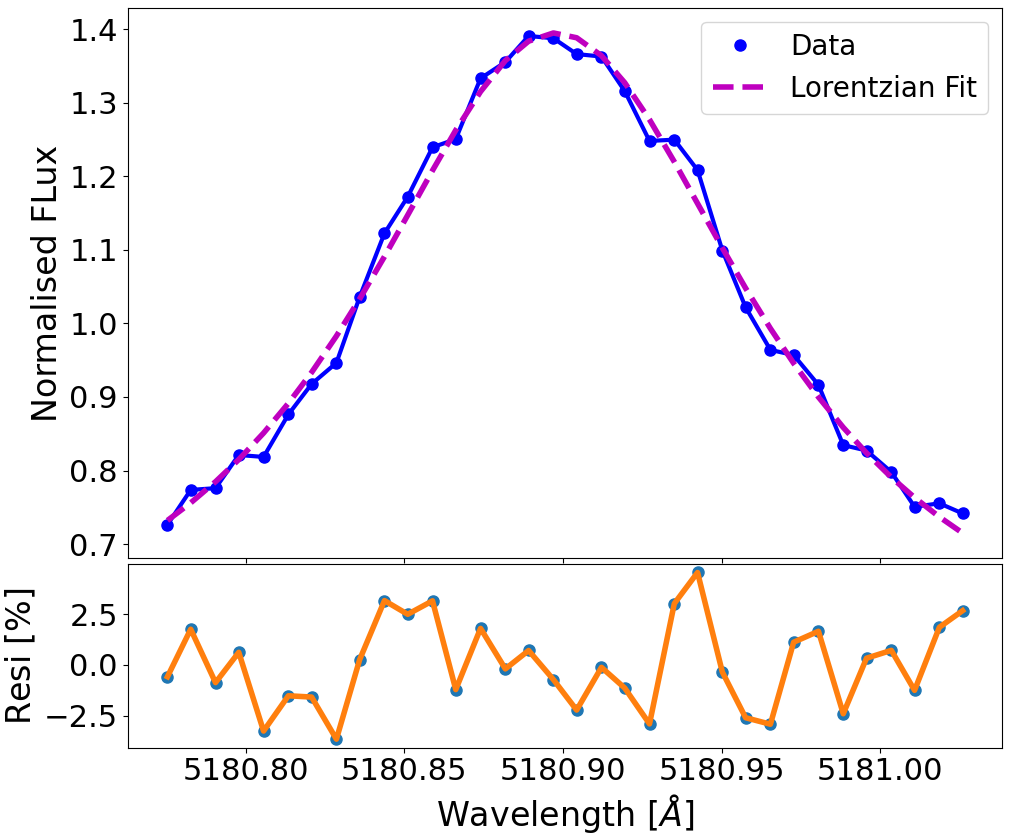}
        \includegraphics[width=2.2in, height=2.2in]{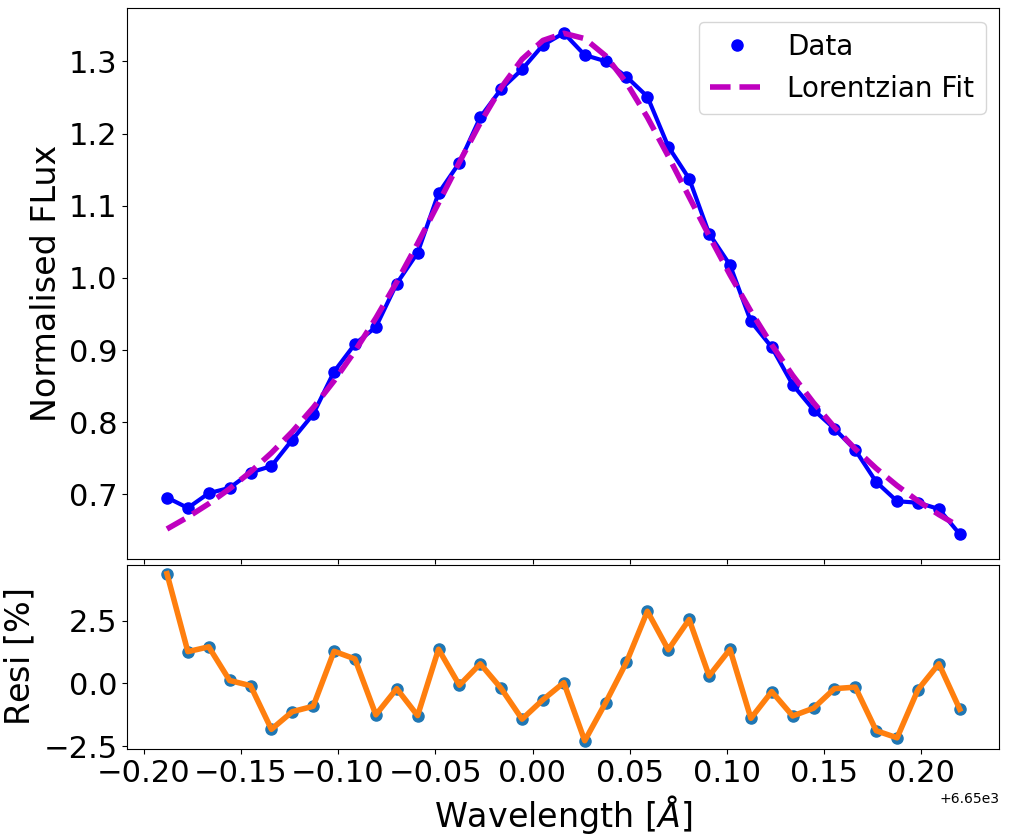} 
        \includegraphics[width=2.2in, height=2.2in]{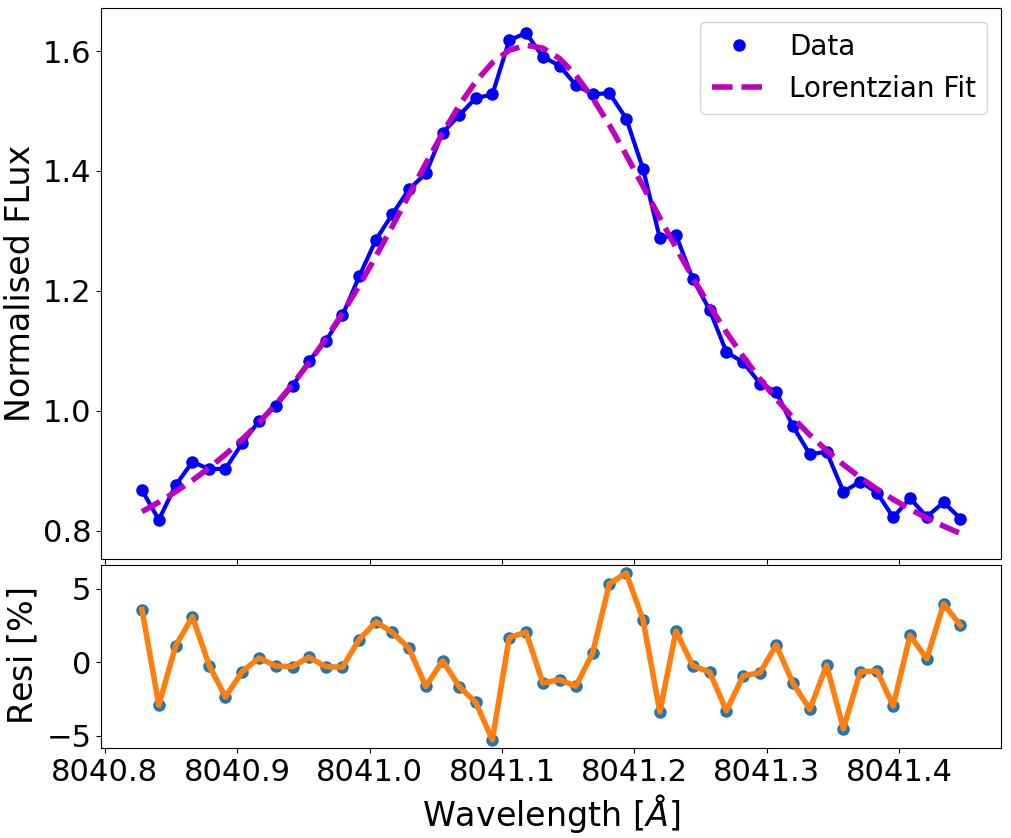} 
	\caption{Lorentzian fits for three transmission peaks corresponding to three resonant modes obtained from three cross-dispersed orders for thin FPM. The central wavelengths and FWHMs of these modes from left to right are 5180.9 $\pm$ 0.016 \AA~ and 0.154 $\pm$ 0.007 \AA, 6650.02 $\pm$ 0.001 \AA~ and 0.215 $\pm$ 0.006 \AA~, and 8041.12 $\pm$ 0.001 \AA~ and 0.312 $\pm$ 0.010 \AA), respectively. The lower plots show the residuals between data and fit.}
	\label{Fig:Gauss_fit}
\end{figure*}

\section{Results}
\subsection{Comparison between Theoretical and Experimental Performances}

\subsubsection{Free Spectral Range Measurements}\label{sec: FreeSpectralRangeMeasurements}
The theoretical FSR was estimated using the eq. \ref{eq:FSR_equation}. For the experimental measurement of FSR, we used an extracted spectrum of FPM at five different temperatures. 
Both Gaussian (for example, \citealt{2021AJ....161..252T}) and Lorentzian functions (for example, \citealt{2015A&A...581A.117B}) have been used for fitting etalon's spectral responses. To derive the central wavelength and FWHM of each FP resonant mode, we used a Lorentzian function with a constant offset based on the spectral response of the FPE, resolution of the spectrograph and FSRs of our FPMs. We choose twice the FWHM window size around each transmission peak to fit the data. To set the initial parameters for the fit, the wavelength-dependent FWHM was predicted from theoretical $FSR$ (using Eq. \ref{eq:FSR_equation}) and from the quoted finesse of the FPM in the catalogue (see Table \ref{tab:MetalonDetailed}). In addition, peak-finding algorithm from SciPy was implemented to identify initial transmission peaks. The fit for three transmission peaks taken from three different orders, for example, is shown in Fig. \ref{Fig:Gauss_fit}. The central wavelength and wavelength dependent FWHM of these modes from the fit are (5180.9 $\pm$ 0.016 \AA~ and 0.154 $\pm$ 0.007), (6650.02 $\pm$ 0.001 \AA~ and 0.215 $\pm$ 0.006,) and (8041.12 $\pm$ 0.001 \AA~ and 0.312 $\pm$ 0.010), respectively.

For each cross-dispersed order, the centres of all transmission peaks were calculated, and the separation between two adjacent peaks for all peaks was estimated. The mean of all separations provides the FSR of the order, and the mid-wavelength of that order was considered as the central wavelength of that particular order. Following this method, the FSRs and central wavelengths of all orders were evaluated. The FSRs versus central wavelengths for thin and thick FPMs are shown in Fig. \ref{Fig:FSR_comparison}. This measurement was carried out for all five temperatures. 

We found that the SNR of extracted FPM spectra is sufficient to identify peaks uniquely and accurately from 5000 (5600) \AA~ to 8000 (8000) \AA~ for thin (thick) FPM, respectively. The theoretical and measured FSRs are matched well for thick FPM in the wavelength range between 5600 \AA~ and 8000 \AA~. However, for thin FPM, the theoretical curve does not match well with the measured FSRs. The reason for such behaviour of the thin FPM is a subject to further investigation and is out of the scope of this paper. However, this could be due to several reasons. For example, we paid insufficient attention to the alignment of FPM and fibre coupling and noting that our theoretical derivation is based on a simple consideration of normal light incidence. 

\subsubsection{Temperature Sensitivity Measurements} \label{sec: TemperatureSensitivityMeasurements}
To understand how much temperature precision is required for the FPE box to get to 1 m/s calibration accuracy, we measured the temperature sensitivity of FPM cavity material (i.e. fused silica). The theoretical drift was measured using the eq. \ref{eq:shift_derivation_final}. In addition, we measured the drift for 2 $^{\circ}$C change in temperature from 20 $^{\circ}$C to 28 $^{\circ}$C as shown in Fig. \ref{Fig:drift_vs_wavelength_for_temperature_change}. We found that there is a significant deviation between theoretical and experimental measurements for both thin and thick FPMs. On one hand, our theoretical approach and on the other hand, our thermal enclosure for Metalon v1.0 are simplistic models to prove our concept. In addition, a detailed investigation of the behaviour of fused silica at different temperatures of interest is required. Furthermore, the optimum fibre coupling, sophisticated alignment method ,and repeatability of our experiment are our prime focus. Nonetheless, Fig. \ref{Fig:drift_vs_wavelength_for_temperature_change} showed that the experimentally measured drifts are lesser than the theoretical prediction. Hence, the theoretical measurement set the lower limit of temperature stability required, whereas the experimental measurement set the upper limit. The drift of 0.039 \AA$/^{\circ}$C was evaluated at 6550 \AA~ from the theoretical derivation of thin FPM. This drift indicated that the temperature of the FPM would have to be controlled to $\pm$0.5 mK. The experimental measurements corresponding to 20 $^{\circ}$C vs 22 $^{\circ}$C displayed the minimal drift. The value is about 0.029 \AA$/^{\circ}$C at 6550 \AA~ which translates to 1.5 mK stability requirement. Therefore, theoretical and experimental evaluation indicated that we have to control the FPM in between 0.5 mK and 1.5 mK for 1 m/s calibration accuracy. This range of required stability is wavelength-independent. We further performed the RV analysis of FPM in comparison with ThAr for both Metalon v1.0 and v2.0.

\subsection{FPM RV Performance} \label{sec: MetalonRVPerformance}
To test and compare the RV performance of the thin FPM with ThAr, we carried out our investigation for both versions of the metalon. As described above, we obtained data over 24 hours and 12 hours for Metalon v1.0 and v2.0, respectively.

\begin{figure*}
	\centering
	\includegraphics[width=3.5in, height=4.2in]{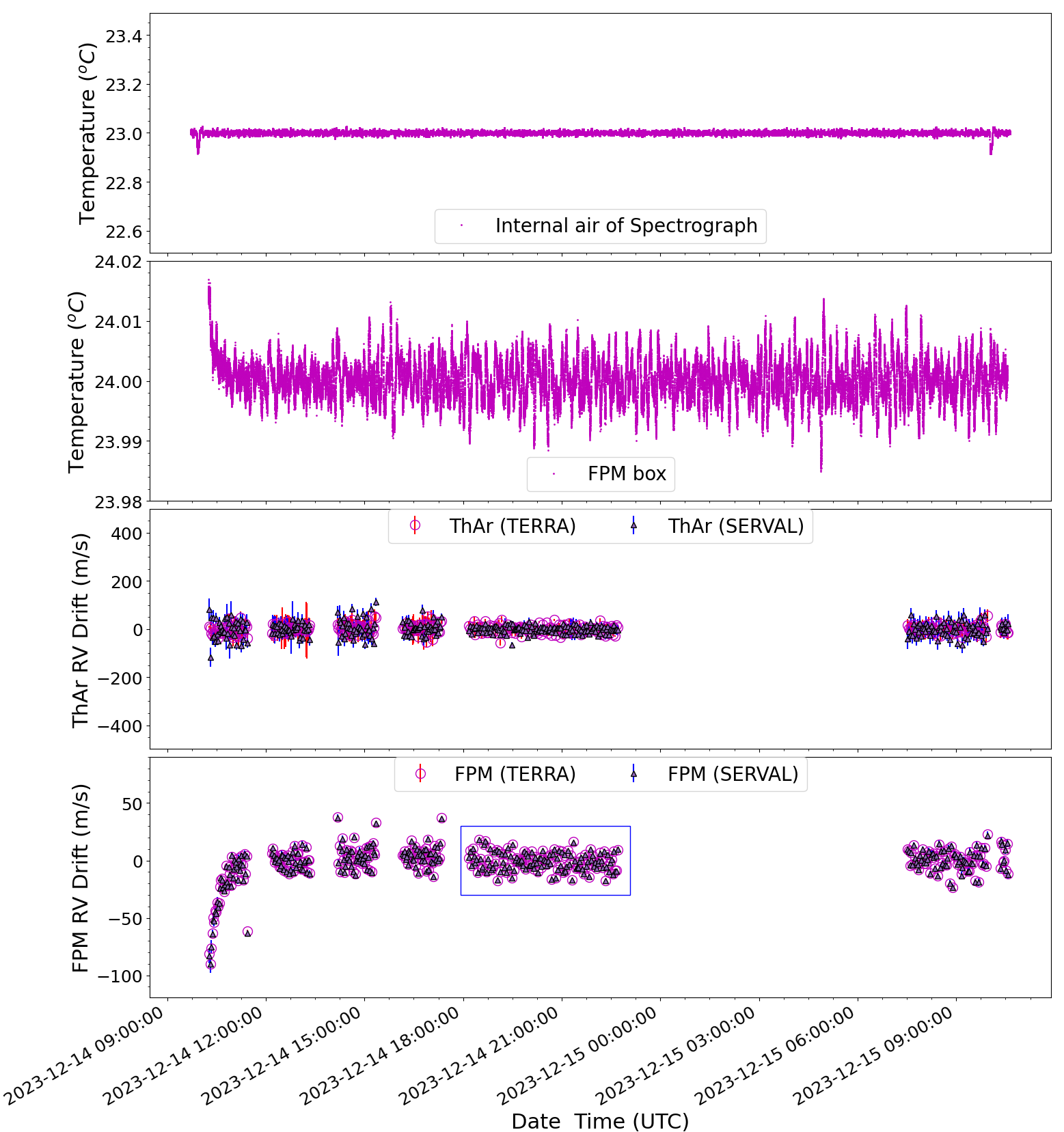}
        \includegraphics[width=3.5in, height=4.2in]{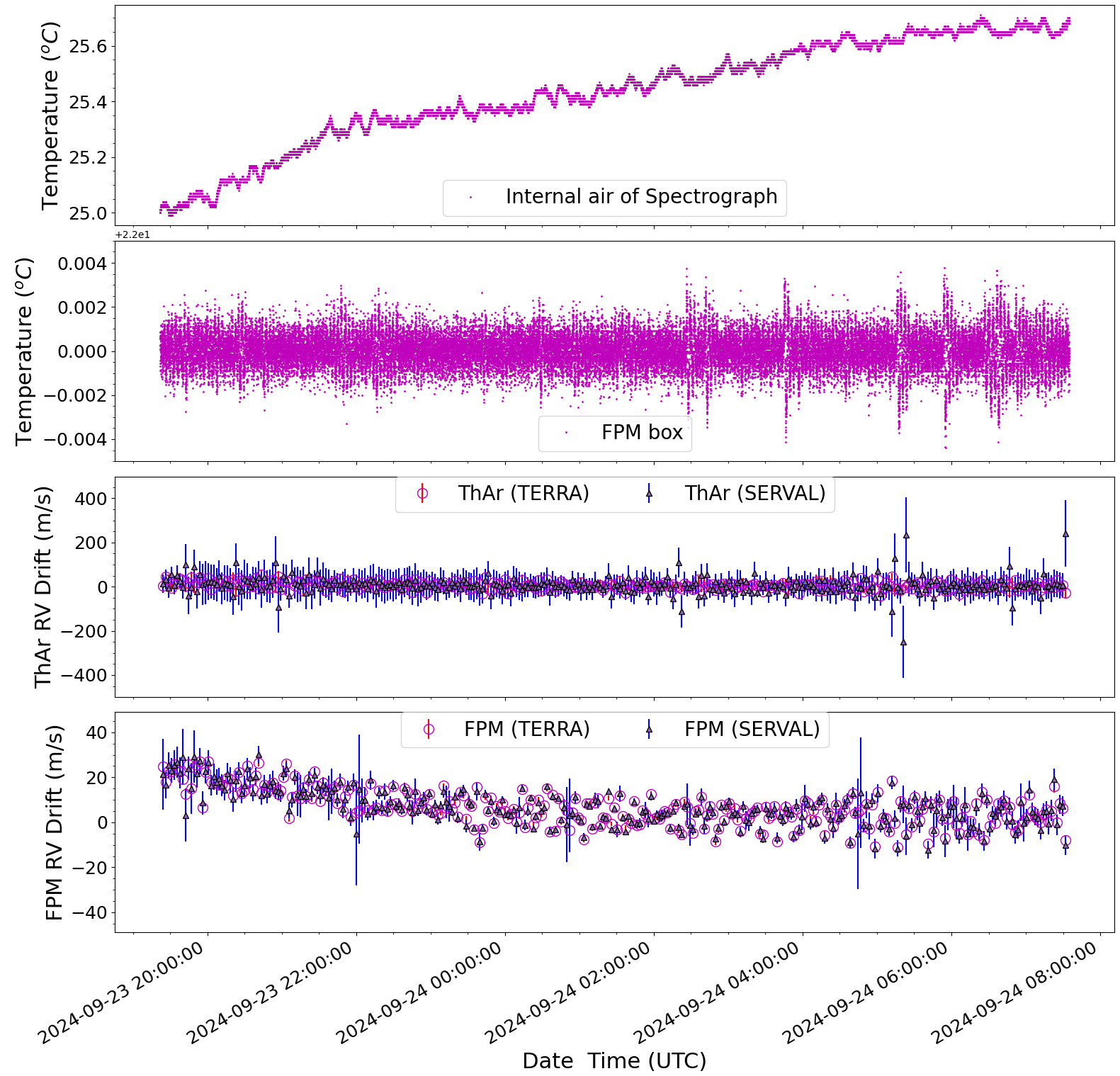}
	\caption{Monitoring the stability for Metalon v1.0 (left) and Metalon v2.0 (right). From top to bottom in each side: variation of internal air temperature of the spectrograph, temperature of FPM box, RV drift of ThAr and RV drift of thin FPM, respectively. We set the y-axis range equal for comparison.}
 	\label{Fig:temperature_RVdrift_versus_time}
\end{figure*}

For Metalon v1.0, our result is presented in Table~\ref{tab:rv_drift_stat} and displayed in Fig. \ref{Fig:temperature_RVdrift_versus_time} (left). Table~\ref{tab:rv_drift_stat} also lists the stability of the internal air temperature of the spectrograph and FPM box temperature. The median and standard deviation of temperature both for spectrograph as well FPM box are 23.0 $^{\circ}$C and 8 m$^{\circ}$C, and 24.0 $^{\circ}$C and 3 m$^{\circ}$C, respectively throughout data acquisition. Considering all data, the RV drift is evaluated with a standard deviation of $>$ 40 m/s for both ThAr and FPM, directly corresponding to the stability of the FPM system as well as the spectrograph. Furthermore, the initial large change in RV drift for FPM was in agreement with the change in temperature of the FPM system. It is to note that, as stated above, we began the data acquisition immediately after setting the temperature to the FPM system. The FPM system took almost an hour to stabilize against its set temperature, and the required time depends on the time constant of the feedback loop and optimum tuning parameters. Visualizing the behaviour of the RV drift over time, the 5$^{th}$ dataset (from 2023-12-14 18:00:00 to 2023-12-14 23:00:00, highlighted with a rectangular box) corresponded to the minimal RV drift. The values are 17 m/s (TERRA) and 15 m/s (SERVAL) for ThAr and 8 m/s (TERRA) and 8 m/s (SERVAL) for FPM.  Our result showed that FPM is superior to ThAr in drift measurement.

\begin{figure*}
	\centering
	\includegraphics[width=3.55in, height=2.8in]{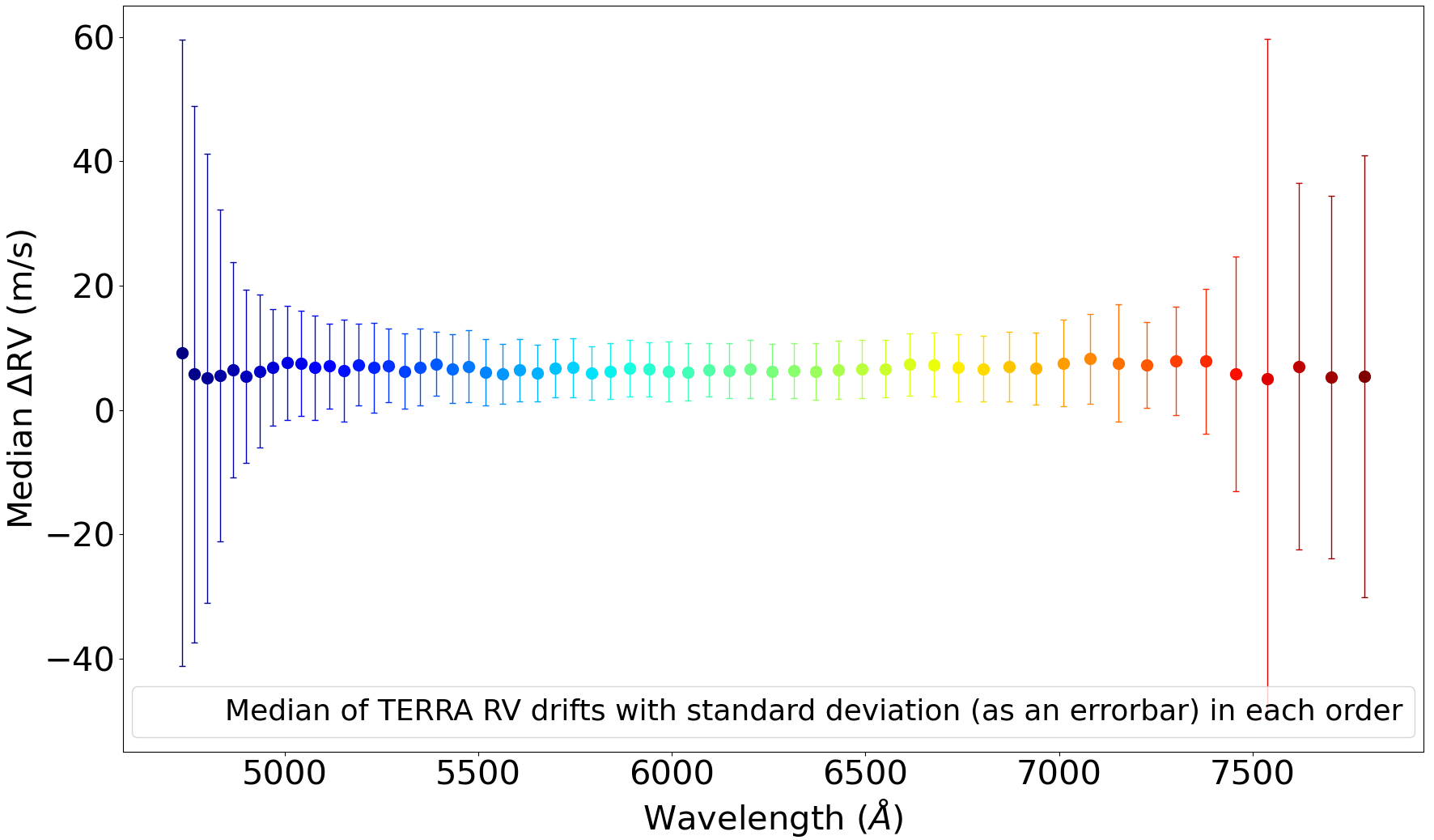}
    \includegraphics[width=3.45in, height=2.8in]{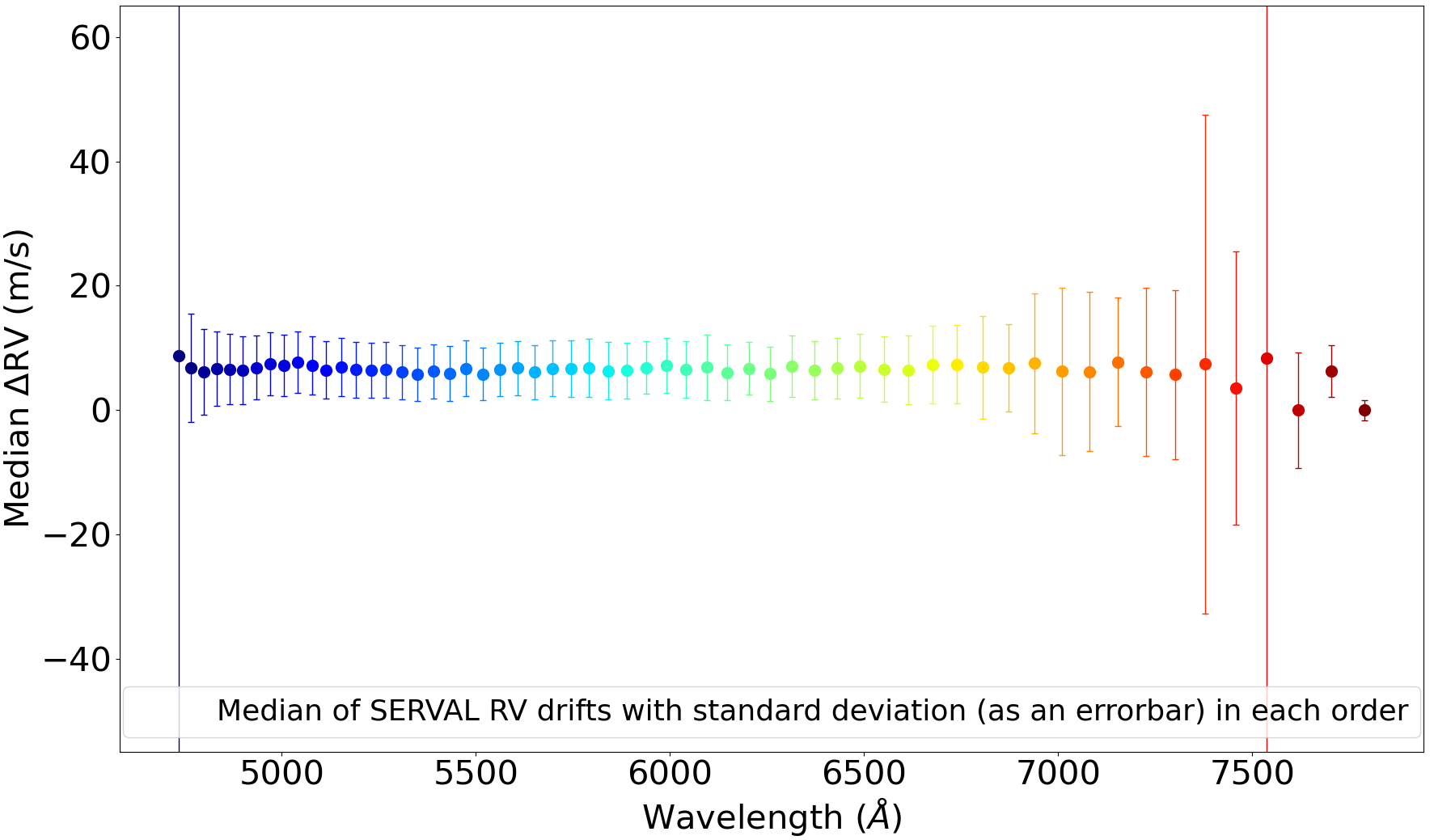}
        \caption{Median RV drift on order-by-order basis for all observation for Metalon v2.0 measured using TERRA (left) and SERVAL (right). They are plotted on the same scale for comparison.}
 	\label{Fig:order-by-order_RVdrift_versus_time}
\end{figure*}

Our experimental result for Metalon v2.0 is listed in Table~\ref{tab:rv_drift_stat} and displayed in Fig. \ref{Fig:temperature_RVdrift_versus_time} (right). The internal air temperature of the spectrograph was gradually raised during our experiment. It was not unusual because the air conditioning was inactive and the spectrograph was not a closed system. Although an active stabilization using a feedback loop was being operated to stabilize the spectrograph, it was certainly not capable of stabilizing the whole laboratory. On the contrary, the FPM system was being stabilized within 0.8 mK. The measured RV drifts for ThAr were 17 m/s and 37 m/s by TERRA and SERVAL, respectively. This difference on RV drift could be believed to be intrinsic to the measurements of TERRA and SERVAL because when the system was not fully stable, consideration of using all spectra to build the template spectrum might be the potential pitfall. Moreover, due to the irregular line distribution and large intensity difference, the SNR of the continuum of the ThAr spectrum depends on instrumental profile and data processing technique. For  FPM, both TERRA and SERVAL yielded RV drifts of 8 m/s. The behaviour of FPM RV drift over time showed a temperature response in the measurements. If we fitted a polynomial to the measurements of FPM RV drift and flattened them, the RV drift would have come down to 5 m/s in the wavelength regime from 4700 \AA~ to 7800 \AA~. Our result confirmed that the FPM is more efficient than ThAr for the simultaneous drift measurement. Furthermore, we studied RV drift for each cross-dispersed order for both SERVAL and TERRA, as shown in Fig. \ref{Fig:order-by-order_RVdrift_versus_time}. The median and standard deviation of RV drift from all observations for all orders have been estimated, and the median wavelength in each order provided the representative wavelength of each order. We found that the precision of the RV drift is almost consistent over the wavelength range between 5000 \AA~ and 7000 \AA~ and declined at the two ends of the spectral coverage. It is expected as the SNR reduces toward both ends of the spectral coverage. The SNR of the FPM spectrum depends on grating efficiency, throughput of the spectrograph,
selection of broadband source, alignment of FPM, as well as spectrograph systems and fibre coupling. 

\begin{table}
	 \centering
	\caption{Statistics for RV drifts.}
	\label{tab:rv_drift_stat}
         \resizebox{0.46\textwidth}{!}{%
	   \begin{tabular}{lccl}
          \hline
          \multicolumn{4}{|c|}{Metalon v1.0} \\
           \hline
           Parameters & Median & Standard deviation & Remarks \\
           \hline
           Internal air temperature & 23$^{\circ}$C  & 8 m$^{\circ}$C & 1\\
           of Spectrograph          &                &               & \\
           FPM box temperature      &  24$^{\circ}$C & 4 m$^{\circ}$C & 1\\
           RV drift of ThAr (TERRA) & & 44 m/s & 1\\
           RV drift of ThAr (SERVAL) & & 46 m/s & 1\\           
           RV drift of FPM (TERRA)  & & 47 m/s & 1\\
           RV drift of FPM (SERVAL) & & 47 m/s & 1\\
                                    & &        & \\
           Internal air temperature & 23$^{\circ}$C  & 8 m$^{\circ}$C & 2\\
           of Spectrograph          &                &               & \\
           FPM box temperature      &  24$^{\circ}$C & 4 m$^{\circ}$C & 2\\
           RV drift of ThAr (TERRA) & & 17 m/s & 2\\
           RV drift of ThAr (SERVAL) & & 15 m/s & 2\\           
           RV drift of FPM (TERRA)  & & 8 m/s & 2\\
           RV drift of FPM (SERVAL) & & 8 m/s & 2\\
           \hline
           \multicolumn{4}{|c|}{Metalon v2.0} \\
            \hline
           Internal air temperature & 25.4$^{\circ}$C  & 0.19 $^{\circ}$C & 1\\
           of Spectrograph          &                &               & \\
           FPM box temperature      &  22$^{\circ}$C & 0.8 m$^{\circ}$C & 1\\
           RV drift of ThAr (TERRA) & & 17 m/s & 1\\
           RV drift of ThAr (SERVAL) & & 37 m/s & 1\\           
           RV drift of FPM (TERRA)  & & 8 m/s & 1\\
           RV drift of FPM (SERVAL) & & 8 m/s & 1\\
           \hline
           Notes: (1) - All data; \\ (2) A subset of data (see text) 
           \end{tabular}}
\end{table}

\section{Discussion} \label{sec:discussion}
In this work, we have presented preliminary tests of a commercially available FPM. Commercial off-the-shelf components have been used for parts of the FPM system with a consideration of the cost of development. The finesse of such a plane parallel FPM is limited as the diffraction losses are quite large and thereby decreasing the reflection coefficient.

We have achieved a temperature stability of 0.8 mK with precise thermal control of the FPM system by using a dewar flask and active stabilisation. We have estimated that such a temperature stability can provide a RV stability of about 1 m s$^{-1}$. However, the temporal drift of the FPM calibrator was measured as 8 m s$^{-1}$, which is below our requirement for EXOhSPEC. Nonetheless, a low-cost spectrograph with a typical accuracy of 10 m s$^{-1}$ does enable several potential science cases, for example,  the detection and characterization of hot Jupiters like exoplanets and young brown dwarfs, the study of binary star systems and stellar oscillations in red giants.

We believe that the temporal drift value in this work arose because when these results were taken the spectrograph active stabilisation method (discussed in \citealt{2021PASP..133b5001J}) were not engaged. Probably, the precision of the obtained FPM stability was influenced by several external factors, which need to be carefully controlled for further improvement. For example, a marginal opto-mechanical misalignment of fibre-fibre connectors and the FPM system among the components that includes the illuminating input fiber, two parallel mirrors, and the output fibre, can change the illumination of the spectrograph significantly. This change in illumination further causes line broadening, shifts the centroid of the spectral line, results in effective finesse loss and introduces error in the measured radial velocity \citep{10.1117/12.857951, Schwab_2015}. A misalignment in the order of a few microns can cause an error in the order of meters per second on recorded radial velocity (\citealt{10.1117/12.926232}). A robust and sophisticated alignment method is essential to suppress such misalignment errors. These errors can be addressed by regular comparison between the FPM lines and spectral lines of known absolute frequency, as performed on HARPS against a ThAr lamp every night. Furthermore, we used a bifurcated fibre to feed the spectrograph for our experiment. The two breakout legs of bifurcated fibre should be aligned horizontally to the optic axis at the entrance of the spectrograph to form two nearby simultaneous spectra for both fibres on the detector. In optimum alignment, a trace of the calibration fibre should fall exactly in the middle of two cross-dispersed orders of the science fibre. However, as can be seen from Fig. \ref{Fig:metalon-spectrum_comparison}, it is apparent that two breakout legs of the bifurcated fibre are not optimally aligned. As a result, the ThAr continuum was affected by the FPM continuum. Thus, we believe that the result can be improved further if the bifurcated fibre is aligned optimally.

In addition, a perfectly aligned system would be less sensitive to the fibre modal noise. The modal noise, which causes the change in the illumination pattern of the spectrograph, becomes increasingly more problematic as the number of modes propagating through the fibre decreases. In our investigation, we have used fibres of core 62.5 $\mu$m and 10 $\mu$m. A 10 $\mu$m fibre would show more modal noise effect than a 62.5 $\mu$m fibre (see \citealt{10.1093/rasti/rzad059} and references therein). \citet{10.1093/rasti/rzad059} also found that a 10 $\mu$m fibre is very responsive to thermal and mechanical changes. However, we carried out our experiment without implementing any modal noise mitigation technique, which might impact our result.

The spectral SNR was much better for the thin FPM than the thick FPM for an exposure of 100 s (see, Fig. \ref{Fig:extracted_spectrum}). The measured transmission curves of the FPM systems (see, Fig. \ref{Fig:FPM_Transmission}) confirms that the better spectral SNR is obtainable for thin FPM in comparison to thick FPM for a given exposure. The SNR of the thick FPM can be optimised either by setting up an optimised exposure time or increasing the output power of the broadband source. An increase in exposure time saturates several ThAr lines, influencing the wavelength solution, hence the data reduction and experimental result. In addition, enhancement of scattered light into the spectrograph with an increase of broadband power might affect the adjacent cross-dispersed orders separation during the data processing. Moreover, SNR is limited by the finesse of FPM and throughput of the spectrograph. Furthermore, Fig. \ref{Fig:extracted_spectrum} displays an envelope of the FPM spectrum, which is relatively strong in the thick FPM. The most likely cause is optical beating in the reflecting and transmitting spectra of multibeam interference in the FPM resonant cavity. Due to this envelope, specifically at minimums, care should be taken in analysing spectrum, especially for thick FPM to avoid any algorithmic issues with the misidentification of minimums and maximums. 

In the design and manufacture of a solid FPM, fused Silica is used, which is believed to be chemically extremely pure, quite inert, homogeneous and optically isotropic. However, such material is sensitive to two forms of temperature effects as mentioned above. Fused Silica has a positive temperature coefficient, implying the increase of refractive index with increasing temperature. In addition, the physical thickness of the cavity material i.e. the thickness of the FPM itself goes up as temperature increases, implying a positive coefficient of thermal expansion. The control that we have to minimize these temperature effects is to control the temperature of the FPM system. Now, we can determine the cavity thickness at our achieved temperature stability. For an accuracy of 1 m/s, the cavity thickness needs to be known better than $\frac{\partial{\lambda}}{\lambda} = 3\times10^{-9}$. Following \citet{2019A&A...624A.122C} (see, sec 2), the measured cavity thickness variations for one peak at each of three different wavelengths for Metalon v2.0 are 5.9$\times$10$^{-5}$ \AA~, 7.4$\times$10$^{-5}$ \AA~ and 3.7$\times$10$^{-5}$ \AA~ at about 5000 \AA, 6000 \AA~ and 7500 \AA, translating into $\frac{\partial{\lambda}}{\lambda}$ as 1.2$\times$10$^{-8}$, 1.2$\times$10$^{-8}$ and 0.5$\times$10$^{-8}$, respectively. This implies that the cavity spacing can be derived with an accuracy of better than 4 m s$^{-1}$. However, the spectrograph was not stable at the level of accuracy we would like to during this experiment. In addition, we computed finesses at three wavelengths from our measured FWHMs and FSRs for thin FPM (see, sec \ref{sec: FreeSpectralRangeMeasurements}). A finesse of 1.6 at about 5200 \AA~, 2.1 at about 6650 \AA~ and 2.1 at about 8050 \AA~ was evaluated for thin FPM. As the FPM lines are convolved with the instrument profile, corresponding to the spectrograph resolution, a lower value would be expected than that provided by the manufacturer (i.e., 2.5). Further work will be necessary to determine cavity spacing for each peak and finesse as a function of wavelength. Moreover, this solid FPM is metal coated for enhancing the finesse. At this point, we have no information about the long-term behaviour of the mirror coating and thus, long-term monitoring on a highly stabilized spectrograph is required to understand the nature and performance of a metalon system for the wavelength calibration of a high resolution spectrograph.

\section{Summary}
In this paper, we have demonstrated the development of low-cost solid fused silica etalons with broadband metallic coatings for wavelength calibration for high-resolution radial velocity spectrographs. The off-the-shelf components have been used for the development of the metalon system. We characterised the metalon and investigated its thermal performance. As this is a solid-cavity metalon, the only parameter that needs to be controlled for environmental changes is temperature. 

The temperature of the FPM system was controlled with active temperature methodology by using a controlled loop feedback system. For investigating the thermal performance of the fused silica, the set-point temperature of the FPM system was varied from 20 $^{\circ}$C to 28 $^{\circ}$C in steps of 2 $^{\circ}$C and a high-resolution spectrum obtained at each temperature using a broadband continuum source (tungsten lamp). We derived the free spectral range for the metalon for each temperature and found that our experimental measurements corroborated with the theoretical derivation. We further measured the drift in 2$^{\circ}$C temperature change between 20 $^{\circ}$C and 28 $^{\circ}$C. We found that our measurements deviated significantly from the theoretical measurement. There could be various reasons for this deviation, for example, the precision of temperature stability in the FPM system, fibre-coupling, and alignment of the FPM mirrors. In addition, our theoretical derivation is based on a simplistic approach with a consideration of the normal incidence of light. Nonetheless, this comparison shed light on the thermal response of silica with wavelength and showed that the temperature of the system needed to be controlled in between 0.5 mK and 1.5 mK to achieve a RV stability of 1 m/s. 

The use of a dewar flask with active stabilisation demonstrated the potential to keep the temperature ﬂuctuations to a short-term ($\sim$ 12 hrs) standard deviation of 0.8 mK. Using TERRA and SERVAL standard radial velocity packages and by considering the FPM system as a Sun for barycentric correction, our RV analysis for FPM showed a temporal drift of 8 m/s in the wavelength regime from 4700 \AA~ to 7800 \AA~. We noted that the precision of RV measurements diminished at two ends of the wavelength coverage. The declining SNR of the FPM spectrum at two ends could be the reason. Long-term monitoring of our FPM is required to understand the detailed characteristics of metalon, however, our result shows promise to get to 1 m/s wavelength accuracy using such calibrators.

\section*{Data Availability}
The data underlying this article will be shared upon reasonable request to the corresponding author.

\section*{Acknowledgements}
Support for this research was provided from the STFC to the University of Hertfordshire through grants for high performance computing infrastructure ST/R000905/1 and experimental equipment provided via STFC grants ST/P005667/1, ST/R006598/1 and ST/T007311/1. We are also grateful to Eugen Guritanu for design and printing support of both Metalon v1.0 and v2.0. We are very grateful to both reviewers for their critical and thoughtful feedback, which helped us to improve the manuscript greatly. We thank Calvin Joy from LightMachinery for their insights on metalon coating. We also thank Kathryn Hogan-Lewis from Photron Pty Ltd for providing information on the availability of ThAr lamps in the market.




\bibliographystyle{rasti}
\bibliography{reference}

\begin{thebibliography}{54}
\expandafter\ifx\csname natexlab\endcsname\relax\def\natexlab#1{#1}\fi

\bibitem[Anglada-Escudé \& Butler(2012)]{Anglada-Escude_2012}
Anglada-Escudé, G. \& Butler, R.~P., 2012.
\newblock The harps-terra project. i. description of the algorithms,
  performance, and new measurements on a few remarkable stars observed by
  harps, {\it The Astrophysical Journal Supplement Series\/}, {\bf 200}(2), 15.

\bibitem[{Baranne} et~al.(1996){Baranne}, {Queloz}, {Mayor}, {Adrianzyk},
  {Knispel}, {Kohler}, {Lacroix}, {Meunier}, {Rimbaud}, \&
  {Vin}]{1996A&AS..119..373B}
{Baranne}, A., {Queloz}, D., {Mayor}, M., {Adrianzyk}, G., {Knispel}, G.,
  {Kohler}, D., {Lacroix}, D., {Meunier}, J.~P., {Rimbaud}, G., \& {Vin}, A.,
  1996.
\newblock {ELODIE: A spectrograph for accurate radial velocity measurements.},
  {\it \aaps\/}, {\bf 119}, 373--390.

\bibitem[{Bauer} et~al.(2015){Bauer}, {Zechmeister}, \&
  {Reiners}]{2015A&A...581A.117B}
{Bauer}, F.~F., {Zechmeister}, M., \& {Reiners}, A., 2015.
\newblock {Calibrating echelle spectrographs with Fabry-P{\'e}rot etalons},
  {\it \aap\/}, {\bf 581}, A117.

\bibitem[{Beckers}(1976)]{1976Natur.260..227B}
{Beckers}, J.~M., 1976.
\newblock {Reliability of sunspots as tracers of solar surface rotation}, {\it
  \nat\/}, {\bf 260}(5548), 227--229.

\bibitem[{Burenkov} et~al.(1979){Burenkov}, {Voikhanskaia}, \&
  {Rylov}]{1979AISAO..11...70B}
{Burenkov}, A.~N., {Voikhanskaia}, N.~F., \& {Rylov}, V.~S., 1979.
\newblock {Atlas of the thorium spectrum.}, {\it Astrofizicheskie Issledovaniia
  Izvestiya Spetsial'noj Astrofizicheskoj Observatorii\/}, {\bf 11}, 70--106.

\bibitem[{Cersullo} et~al.(2017){Cersullo}, {Wildi}, {Chazelas}, \&
  {Pepe}]{2017A&A...601A.102C}
{Cersullo}, F., {Wildi}, F., {Chazelas}, B., \& {Pepe}, F., 2017.
\newblock {A new infrared Fabry-P{\'e}rot-based radial-velocity-reference
  module for the SPIRou radial-velocity spectrograph}, {\it \aap\/}, {\bf 601},
  A102.

\bibitem[{Cersullo} et~al.(2019){Cersullo}, {Coffinet}, {Chazelas}, {Lovis}, \&
  {Pepe}]{2019A&A...624A.122C}
{Cersullo}, F., {Coffinet}, A., {Chazelas}, B., {Lovis}, C., \& {Pepe}, F.,
  2019.
\newblock {New wavelength calibration for echelle spectrographs using
  Fabry-P{\'e}rot etalons}, {\it \aap\/}, {\bf 624}, A122.

\bibitem[Das et~al.(2018)Das, Banyal, Kathiravan, Sivarani, \&
  B.]{10.1117/12.2312472}
Das, T., Banyal, R.~K., Kathiravan, S., Sivarani, T., \& B., R., 2018.
\newblock {Development of a stabilized Fabry-Perot based wavelength calibrator
  for precision Doppler spectroscopy}, in {\em Ground-based and Airborne
  Instrumentation for Astronomy VII\/}, vol. 10702, p. 107026A, International
  Society for Optics and Photonics, SPIE.

\bibitem[Diddams et~al.(2020)Diddams, Vahala, \&
  Udem]{doi:10.1126/science.aay3676}
Diddams, S.~A., Vahala, K., \& Udem, T., 2020.
\newblock Optical frequency combs: Coherently uniting the electromagnetic
  spectrum, {\it Science\/}, {\bf 369}(6501), eaay3676.

\bibitem[{Errmann} et~al.(2020){Errmann}, {Cook}, {Anglada-Escud{\'e}},
  {Sithajan}, {Mkrtichian}, {Semenko}, {Martin}, {Tanvir}, {Feng}, {Collett},
  \& {Jones}]{2020PASP..132f4504E}
{Errmann}, R., {Cook}, N., {Anglada-Escud{\'e}}, G., {Sithajan}, S.,
  {Mkrtichian}, D., {Semenko}, E., {Martin}, W., {Tanvir}, T.~S., {Feng}, F.,
  {Collett}, J.~L., \& {Jones}, H. R.~A., 2020.
\newblock {HiFLEx{\textemdash}A Highly Flexible Package to Reduce
  Cross-dispersed Echelle Spectra}, {\it \pasp\/}, {\bf 132}(1012), 064504.

\bibitem[{Faria, J. P.} et~al.(2022){Faria, J. P.}, {Suárez Mascareño, A.},
  {Figueira, P.}, {Silva, A. M.}, {Damasso, M.}, {Demangeon, O.}, {Pepe, F.},
  {Santos, N. C.}, {Rebolo, R.}, {Cristiani, S.}, {Adibekyan, V.}, {Alibert,
  Y.}, {Allart, R.}, {Barros, S. C. C.}, {Cabral, A.}, {D’Odorico, V.}, {Di
  Marcantonio, P.}, {Dumusque, X.}, {Ehrenreich, D.}, {González Hernández, J.
  I.}, {Hara, N.}, {Lillo-Box, J.}, {Lo Curto, G.}, {Lovis, C.}, {Martins, C.
  J. A. P.}, {Mégevand, D.}, {Mehner, A.}, {Micela, G.}, {Molaro, P.}, {Nunes,
  N. J.}, {Pallé, E.}, {Poretti, E.}, {Sousa, S. G.}, {Sozzetti, A.},
  {Tabernero, H.}, {Udry, S.}, \& {Zapatero Osorio, M. R.}]{Faria2022}
{Faria, J. P.}, {Suárez Mascareño, A.}, {Figueira, P.}, {Silva, A. M.},
  {Damasso, M.}, {Demangeon, O.}, {Pepe, F.}, {Santos, N. C.}, {Rebolo, R.},
  {Cristiani, S.}, {Adibekyan, V.}, {Alibert, Y.}, {Allart, R.}, {Barros, S. C.
  C.}, {Cabral, A.}, {D’Odorico, V.}, {Di Marcantonio, P.}, {Dumusque, X.},
  {Ehrenreich, D.}, {González Hernández, J. I.}, {Hara, N.}, {Lillo-Box, J.},
  {Lo Curto, G.}, {Lovis, C.}, {Martins, C. J. A. P.}, {Mégevand, D.},
  {Mehner, A.}, {Micela, G.}, {Molaro, P.}, {Nunes, N. J.}, {Pallé, E.},
  {Poretti, E.}, {Sousa, S. G.}, {Sozzetti, A.}, {Tabernero, H.}, {Udry, S.},
  \& {Zapatero Osorio, M. R.}, 2022.
\newblock A candidate short-period sub-earth orbiting proxima centauri, {\it
  A\&A\/}, {\bf 658}, A115.

\bibitem[Garoli et~al.(2015)Garoli, Ruffato, Zilio, Calandrini, Angelis,
  Romanato, \& Cattarin]{Garoli:15}
Garoli, D., Ruffato, G., Zilio, P., Calandrini, E., Angelis, F.~D., Romanato,
  F., \& Cattarin, S., 2015.
\newblock Nanoporous gold leaves: preparation, optical characterization and
  plasmonic behavior in the visible and mid-infrared spectral regions, {\it
  Opt. Mater. Express\/}, {\bf 5}(10), 2246--2256.

\bibitem[Ghosh et~al.(2024)Ghosh, Boonsri, Martin, Jones, Choochalerm, Usher,
  Yerolatsitis, Wocial, \& Wright]{10.1093/rasti/rzad059}
Ghosh, S., Boonsri, C., Martin, W., Jones, H. R.~A., Choochalerm, P., Usher,
  S., Yerolatsitis, S., Wocial, T., \& Wright, T., 2024.
\newblock Mitigating modal noise in multimode circular fibres by optical
  agitation using a galvanometer, {\it RAS Techniques and Instruments\/}, {\bf
  3}(1), 8--18.

\bibitem[{Halverson} et~al.(2014){Halverson}, {Mahadevan}, {Ramsey}, {Hearty},
  {Wilson}, {Holtzman}, {Redman}, {Nave}, {Nidever}, {Nelson}, {Venditti},
  {Bizyaev}, \& {Fleming}]{2014PASP..126..445H}
{Halverson}, S., {Mahadevan}, S., {Ramsey}, L., {Hearty}, F., {Wilson}, J.,
  {Holtzman}, J., {Redman}, S., {Nave}, G., {Nidever}, D., {Nelson}, M.,
  {Venditti}, N., {Bizyaev}, D., \& {Fleming}, S., 2014.
\newblock {Development of Fiber Fabry-Perot Interferometers as Stable
  Near-infrared Calibration Sources for High Resolution Spectrographs}, {\it
  \pasp\/}, {\bf 126}(939), 445.

\bibitem[{Hao} et~al.(2021){Hao}, {Tang}, {Ye}, {Hao}, {Han}, {Zhai}, {Zhang},
  {Wei}, \& {Xiao}]{2021AJ....161..258H}
{Hao}, J., {Tang}, L., {Ye}, H., {Hao}, Z., {Han}, J., {Zhai}, Y., {Zhang}, K.,
  {Wei}, R., \& {Xiao}, D., 2021.
\newblock {Effect of Near-field Distribution on Transmission Characteristics of
  Fiber-fed Fabry-Perot Etalons}, {\it \aj\/}, {\bf 161}(6), 258.

\bibitem[Holzwarth et~al.(2000)Holzwarth, Udem, H\"ansch, Knight, Wadsworth, \&
  Russell]{PhysRevLett.85.2264}
Holzwarth, R., Udem, T., H\"ansch, T.~W., Knight, J.~C., Wadsworth, W.~J., \&
  Russell, P. S.~J., 2000.
\newblock Optical frequency synthesizer for precision spectroscopy, {\it Phys.
  Rev. Lett.\/}, {\bf 85}, 2264--2267.

\bibitem[Jennings et~al.(2017)Jennings, Halverson, Terrien, Mahadevan, Ycas, \&
  Diddams]{Jennings:17}
Jennings, J., Halverson, S., Terrien, R., Mahadevan, S., Ycas, G., \& Diddams,
  S.~A., 2017.
\newblock Frequency stability characterization of a broadband fiber
  fabry-p\&\#x000e9;rot interferometer, {\it Opt. Express\/}, {\bf 25}(14),
  15599--15613.

\bibitem[{Jones} et~al.(2021){Jones}, {Martin}, {Anglada-Escud{\'e}},
  {Errmann}, {Campbell}, {Baker}, {Boonsri}, \&
  {Choochalerm}]{2021PASP..133b5001J}
{Jones}, H.~R.~A., {Martin}, W.~E., {Anglada-Escud{\'e}}, G., {Errmann}, R.,
  {Campbell}, D.~A., {Baker}, C., {Boonsri}, C., \& {Choochalerm}, P., 2021.
\newblock {A Small Actively Controlled High-resolution Spectrograph Based on
  Off-the-shelf Components}, {\it \pasp\/}, {\bf 133}(1020), 025001.

\bibitem[Kanodia \& Wright(2018)]{Kanodia_2018}
Kanodia, S. \& Wright, J., 2018.
\newblock Python leap second management and implementation of precise
  barycentric correction (barycorrpy), {\it Research Notes of the AAS\/}, {\bf
  2}(1), 4.

\bibitem[{Kawinkij} et~al.(2019){Kawinkij}, {Prasit}, {Buisset}, {Thummasorn},
  {Kuha}, {Lhospice}, {Jones}, {Martin}, {Errmann}, {Choochalerm},
  {Anglada-Escude}, {Campbell}, {Baker}, {Rattanasoon}, {Aukkaravittayapun},
  {Panyaphirawat}, {Leckngam}, {Mkrtichian}, {Poshyachinda}, \&
  {Soonthornthum}]{2019SPIE11116E..1GK}
{Kawinkij}, A., {Prasit}, A., {Buisset}, C., {Thummasorn}, G., {Kuha}, T.,
  {Lhospice}, E., {Jones}, H. R.~A., {Martin}, W.~E., {Errmann}, R.,
  {Choochalerm}, P., {Anglada-Escude}, G., {Campbell}, D., {Baker}, C.,
  {Rattanasoon}, S., {Aukkaravittayapun}, S., {Panyaphirawat}, T., {Leckngam},
  A., {Mkrtichian}, D., {Poshyachinda}, S., \& {Soonthornthum}, B., 2019.
\newblock {EXOhSPEC collimator mechanical design}, in {\em Astronomical Optics:
  Design, Manufacture, and Test of Space and Ground Systems II\/}, vol. 11116
  of {\bf Society of Photo-Optical Instrumentation Engineers (SPIE) Conference
  Series}, p. 111161G.

\bibitem[{Koch}(1983)]{1983PhDT.........2K}
{Koch}, A., 1983.
\newblock {\it {Spectroscopic determination of solar rotation from Doppler
  measurements of umbral and nearby photospheric plasmas Comparison with tracer
  measurements on sunspots}\/}, Ph.D. thesis, -.

\bibitem[{Koch} \& {Woehl}(1984)]{1984A&A...134..134K}
{Koch}, A. \& {Woehl}, H., 1984.
\newblock {The use of molecular iodine absorption lines as wavelength
  references for solar Doppler shift measurements}, {\it \aap\/}, {\bf 134}(1),
  134--138.

\bibitem[{Kreider} et~al.(2025){Kreider}, {Fredrick}, {Diddams}, {Terrien},
  {Mahadevan}, {Ninan}, {Halverson}, {Bender}, {Hearty}, {Mitchell},
  {Rajagopal}, {Roy}, {Schwab}, \& {Wright}]{2025NatAs.tmp...54K}
{Kreider}, M.~K., {Fredrick}, C., {Diddams}, S.~A., {Terrien}, R.~C.,
  {Mahadevan}, S., {Ninan}, J.~P., {Halverson}, S., {Bender}, C.~F., {Hearty},
  F., {Mitchell}, D., {Rajagopal}, J., {Roy}, A., {Schwab}, C., \& {Wright},
  J.~T., 2025.
\newblock {Quantification of broadband chromatic drifts in Fabry-P{\'e}rot
  resonators for exoplanet science}, {\it Nature Astronomy\/}.

\bibitem[Lee et~al.(2007)Lee, Lee, Jung, Jung, Hwangbo, Kim, \&
  Yoon]{articleLee2007}
Lee, G.~J., Lee, Y., Jung, B.-Y., Jung, S., Hwangbo, C.~K., Kim, J., \& Yoon,
  C., 2007.
\newblock Microstructural and nonlinear optical properties of thin silver films
  near the optical percolation threshold, {\it Journal of Korean Physical
  Society\/}, {\bf 51}, 1555.

\bibitem[Leviton \&
  Frey(2008)]{leviton2008temperaturedependentabsoluterefractiveindex}
Leviton, D.~B. \& Frey, B.~J., 2008.
\newblock Temperature-dependent absolute refractive index measurements of
  synthetic fused silica.

\bibitem[{Lhospice} et~al.(2019){Lhospice}, {Buisset}, {Jones}, {Martin},
  {Errmann}, {Sithajan}, {Boonsri}, {Choochalerm}, {Anglada-Escud{\'e}},
  {Campbell}, {Alagao}, {Paenoi}, {Prasit}, {Panyaphirawat}, {Rattanasoon},
  {Leckngam}, {Mkrtichian}, \& {Aukkaravittayapun}]{2019SPIE11117E..0ZL}
{Lhospice}, E., {Buisset}, C., {Jones}, H. R.~A., {Martin}, W.~E., {Errmann},
  R., {Sithajan}, S., {Boonsri}, C., {Choochalerm}, P., {Anglada-Escud{\'e}},
  G., {Campbell}, D., {Alagao}, M.~A., {Paenoi}, J., {Prasit}, A.,
  {Panyaphirawat}, T., {Rattanasoon}, S., {Leckngam}, A., {Mkrtichian}, D., \&
  {Aukkaravittayapun}, S., 2019.
\newblock {EXOhSPEC folded design optimization and performance estimation}, in
  {\em Society of Photo-Optical Instrumentation Engineers (SPIE) Conference
  Series\/}, vol. 11117 of {\bf Society of Photo-Optical Instrumentation
  Engineers (SPIE) Conference Series}, p. 111170Z.

\bibitem[{Lovis} \& {Pepe}(2007)]{2007A&A...468.1115L}
{Lovis}, C. \& {Pepe}, F., 2007.
\newblock {A new list of thorium and argon spectral lines in the visible}, {\it
  \aap\/}, {\bf 468}(3), 1115--1121.

\bibitem[{Lovis} et~al.(2006){Lovis}, {Pepe}, {Bouchy}, {Lo Curto}, {Mayor},
  {Pasquini}, {Queloz}, {Rupprecht}, {Udry}, \& {Zucker}]{2006SPIE.6269E..0PL}
{Lovis}, C., {Pepe}, F., {Bouchy}, F., {Lo Curto}, G., {Mayor}, M., {Pasquini},
  L., {Queloz}, D., {Rupprecht}, G., {Udry}, S., \& {Zucker}, S., 2006.
\newblock {The exoplanet hunter HARPS: unequalled accuracy and perspectives
  toward 1 cm s $^{-1}$ precision}, in {\em Ground-based and Airborne
  Instrumentation for Astronomy\/}, vol. 6269 of {\bf Society of Photo-Optical
  Instrumentation Engineers (SPIE) Conference Series}, p. 62690P.

\bibitem[Malitson(1965)]{Malitson:65}
Malitson, I.~H., 1965.
\newblock Interspecimen comparison of the refractive index of fused
  silica$\ast$,†, {\it J. Opt. Soc. Am.\/}, {\bf 55}(10), 1205--1209.

\bibitem[{Marcy} \& {Butler}(1992)]{1992PASP..104..270M}
{Marcy}, G.~W. \& {Butler}, R.~P., 1992.
\newblock {Precision Radial Velocities with an Iodine Absorption cell}, {\it
  \pasp\/}, {\bf 104}, 270.

\bibitem[{Mayor} et~al.(2009){Mayor}, {Bonfils}, {Forveille}, {Delfosse},
  {Udry}, {Bertaux}, {Beust}, {Bouchy}, {Lovis}, {Pepe}, {Perrier}, {Queloz},
  \& {Santos}]{2009A&A...507..487M}
{Mayor}, M., {Bonfils}, X., {Forveille}, T., {Delfosse}, X., {Udry}, S.,
  {Bertaux}, J.~L., {Beust}, H., {Bouchy}, F., {Lovis}, C., {Pepe}, F.,
  {Perrier}, C., {Queloz}, D., \& {Santos}, N.~C., 2009.
\newblock {The HARPS search for southern extra-solar planets. XVIII. An
  Earth-mass planet in the GJ 581 planetary system}, {\it \aap\/}, {\bf
  507}(1), 487--494.

\bibitem[{M{\'e}gevand} et~al.(2014){M{\'e}gevand}, {Zerbi}, {Di Marcantonio},
  {Cabral}, {Riva}, {Abreu}, {Pepe}, {Cristiani}, {Rebolo Lopez}, {Santos},
  {Dekker}, {Aliverti}, {Allende Prieto}, {Amate}, {Avila}, {Baldini}, {Bandy},
  {Bristow}, {Broeg}, {Cirami}, {Coelho}, {Conconi}, {Coretti}, {Cupani},
  {D'Odorico}, {De Caprio}, {Delabre}, {Dorn}, {Figueira}, {Fragoso},
  {Galeotta}, {Genolet}, {Gomes}, {Gonz{\'a}lez Hern{\'a}ndez}, {Hughes},
  {Iwert}, {Kerber}, {Landoni}, {Lizon}, {Lovis}, {Maire}, {Mannetta},
  {Martins}, {Molaro}, {Monteiro}, {Moschetti}, {Oliveira}, {Zapatero Osorio},
  {Poretti}, {Rasilla}, {Santana Tschudi}, {Santos}, {Sosnowska}, {Sousa},
  {Tenegi}, {Toso}, {Vanzella}, \& {Viel}]{2014SPIE.9147E..1HM}
{M{\'e}gevand}, D., {Zerbi}, F.~M., {Di Marcantonio}, P., {Cabral}, A., {Riva},
  M., {Abreu}, M., {Pepe}, F., {Cristiani}, S., {Rebolo Lopez}, R., {Santos},
  N.~C., {Dekker}, H., {Aliverti}, M., {Allende Prieto}, C., {Amate}, M.,
  {Avila}, G., {Baldini}, V., {Bandy}, T., {Bristow}, P., {Broeg}, C.,
  {Cirami}, R., {Coelho}, J., {Conconi}, P., {Coretti}, I., {Cupani}, G.,
  {D'Odorico}, V., {De Caprio}, V., {Delabre}, B., {Dorn}, R., {Figueira}, P.,
  {Fragoso}, A., {Galeotta}, S., {Genolet}, L., {Gomes}, R., {Gonz{\'a}lez
  Hern{\'a}ndez}, J., {Hughes}, I., {Iwert}, O., {Kerber}, F., {Landoni}, M.,
  {Lizon}, J.-L., {Lovis}, C., {Maire}, C., {Mannetta}, M., {Martins}, C.
  C.~J.~A.~P., {Molaro}, P., {Monteiro}, M. A.~S., {Moschetti}, M., {Oliveira},
  A., {Zapatero Osorio}, M.~R., {Poretti}, E., {Rasilla}, J.~L., {Santana
  Tschudi}, S., {Santos}, P., {Sosnowska}, D., {Sousa}, S., {Tenegi}, F.,
  {Toso}, G., {Vanzella}, E., \& {Viel}, M., 2014.
\newblock {ESPRESSO: the radial velocity machine for the VLT}, in {\em
  Ground-based and Airborne Instrumentation for Astronomy V\/}, vol. 9147 of
  {\bf Society of Photo-Optical Instrumentation Engineers (SPIE) Conference
  Series}, p. 91471H.

\bibitem[Metcalf et~al.(2019)Metcalf, Fredrick, Terrien, Papp, \&
  Diddams]{Metcalf:19}
Metcalf, A.~J., Fredrick, C.~D., Terrien, R.~C., Papp, S.~B., \& Diddams,
  S.~A., 2019.
\newblock {30 ghz electro-optic frequency comb spanning 300 thz in the
  near infrared and visible}, {\it Opt. Lett.\/}, {\bf 44}(11), 2673--2676.

\bibitem[{Murphy} et~al.(2007){Murphy}, {Udem}, {Holzwarth}, {Sizmann},
  {Pasquini}, {Araujo-Hauck}, {Dekker}, {D'Odorico}, {Fischer}, {H{\"a}nsch},
  \& {Manescau}]{2007MNRAS.380..839M}
{Murphy}, M.~T., {Udem}, T., {Holzwarth}, R., {Sizmann}, A., {Pasquini}, L.,
  {Araujo-Hauck}, C., {Dekker}, H., {D'Odorico}, S., {Fischer}, M.,
  {H{\"a}nsch}, T.~W., \& {Manescau}, A., 2007.
\newblock {High-precision wavelength calibration of astronomical spectrographs
  with laser frequency combs}, {\it \mnras\/}, {\bf 380}(2), 839--847.

\bibitem[{Nave} et~al.(2018){Nave}, {Kerber}, {Den Hartog}, \& {Lo
  Curto}]{2018SPIE10704E..07N}
{Nave}, G., {Kerber}, F., {Den Hartog}, E.~A., \& {Lo Curto}, G., 2018.
\newblock {The dirt in astronomy's genie lamp: ThO contamination of Th-Ar
  calibration lamps}, in {\em Observatory Operations: Strategies, Processes,
  and Systems VII\/}, vol. 10704 of {\bf Society of Photo-Optical
  Instrumentation Engineers (SPIE) Conference Series}, p. 1070407.

\bibitem[{Pepe} et~al.(2000){Pepe}, {Mayor}, {Delabre}, {Kohler}, {Lacroix},
  {Queloz}, {Udry}, {Benz}, {Bertaux}, \& {Sivan}]{2000SPIE.4008..582P}
{Pepe}, F., {Mayor}, M., {Delabre}, B., {Kohler}, D., {Lacroix}, D., {Queloz},
  D., {Udry}, S., {Benz}, W., {Bertaux}, J.-L., \& {Sivan}, J.-P., 2000.
\newblock {HARPS: a new high-resolution spectrograph for the search of
  extrasolar planets}, in {\em Optical and IR Telescope Instrumentation and
  Detectors\/}, vol. 4008 of {\bf Society of Photo-Optical Instrumentation
  Engineers (SPIE) Conference Series}, pp. 582--592.

\bibitem[{Perot} \& {Fabry}(1899)]{1899ApJ.....9...87P}
{Perot}, A. \& {Fabry}, C., 1899.
\newblock {On the Application of Interference Phenomena to the Solution of
  Various Problems of Spectroscopy and Metrology}, {\it \apj\/}, {\bf 9}, 87.

\bibitem[{Reiners} et~al.(2024){Reiners}, {Debus}, {Sch{\"a}fer}, {Tiemann}, \&
  {Zechmeister}]{2024A&A...690A.210R}
{Reiners}, A., {Debus}, M., {Sch{\"a}fer}, S., {Tiemann}, E., \& {Zechmeister},
  M., 2024.
\newblock {Accurate calibration spectra for precision radial velocities: Iodine
  absorption referenced by a laser frequency comb}, {\it \aap\/}, {\bf 690},
  A210.

\bibitem[{Sarmiento, L. F.} et~al.(2018){Sarmiento, L. F.}, {Reiners, A.},
  {Huke, P.}, {Bauer, F. F.}, {Guenter, E. W.}, {Seemann, U.}, \& {Wolter,
  U.}]{Sarmiento_refId0}
{Sarmiento, L. F.}, {Reiners, A.}, {Huke, P.}, {Bauer, F. F.}, {Guenter, E.
  W.}, {Seemann, U.}, \& {Wolter, U.}, 2018.
\newblock Comparing the emission spectra of u and th hollow cathode lamps and a
  new u line list, {\it A\&A\/}, {\bf 618}, A118.

\bibitem[Sch{\"a}fer \& Reiners(2012)]{10.1117/12.926232}
Sch{\"a}fer, S. \& Reiners, A., 2012.
\newblock {Two Fabry-Perot interferometers for high precision wavelength
  calibration in the near-infrared}, in {\em Ground-based and Airborne
  Instrumentation for Astronomy IV\/}, vol. 8446, p. 844694, International
  Society for Optics and Photonics, SPIE.

\bibitem[Schmidt \& Bouchy(2024)]{10.1093/mnras/stae920}
Schmidt, T.~M. \& Bouchy, F., 2024.
\newblock Characterization of the espresso line-spread function and improvement
  of the wavelength calibration accuracy, {\it Monthly Notices of the Royal
  Astronomical Society\/}, {\bf 530}(1), 1252--1273.

\bibitem[{Schmidt} et~al.(2022){Schmidt}, {Chazelas}, {Lovis}, {Dumusque},
  {Bouchy}, {Pepe}, {Figueira}, \& {Sosnowska}]{2022A&A...664A.191S}
{Schmidt}, T.~M., {Chazelas}, B., {Lovis}, C., {Dumusque}, X., {Bouchy}, F.,
  {Pepe}, F., {Figueira}, P., \& {Sosnowska}, D., 2022.
\newblock {Chromatic drift of the Espresso Fabry-P{\'e}rot etalon}, {\it
  \aap\/}, {\bf 664}, A191.

\bibitem[Schmidt et~al.(2025)Schmidt, Reiners, Murphy, Lo Curto, Martins, \&
  Huke]{10.1093/mnras/staf588}
Schmidt, T.~M., Reiners, A., Murphy, M.~T., Lo Curto, G., Martins, C. J.
  A.~P., \& Huke, P., 2025.
\newblock Validation of the espresso wavelength calibration using iodine
  absorption cell spectra, {\it Monthly Notices of the Royal Astronomical
  Society\/}, {\bf 539}(4), 3301--3318.

\bibitem[Schwab et~al.(2015)Schwab, Stürmer, Gurevich, Führer, Lamoreaux,
  Walther, \& Quirrenbach]{Schwab_2015}
Schwab, C., Stürmer, J., Gurevich, Y.~V., Führer, T., Lamoreaux, S.~K.,
  Walther, T., \& Quirrenbach, A., 2015.
\newblock Stabilizing a fabry–perot etalon peak to 3 cm s-1 for
  spectrograph calibration, {\it Publications of the Astronomical Society of
  the Pacific\/}, {\bf 127}(955), 880.

\bibitem[{Sharma} \& {Chakraborty}(2021)]{2021JATIS...7c8005S}
{Sharma}, R. \& {Chakraborty}, A., 2021.
\newblock {Precision wavelength calibration for radial velocity measurements
  using uranium lines between 3800 and 6900 {\r{A}}}, {\it Journal of
  Astronomical Telescopes, Instruments, and Systems\/}, {\bf 7}, 038005.

\bibitem[{Steinmetz} et~al.(2008){Steinmetz}, {Wilken}, {Araujo-Hauck},
  {Holzwarth}, {H{\"a}nsch}, {Pasquini}, {Manescau}, {D'Odorico}, {Murphy},
  {Kentischer}, {Schmidt}, \& {Udem}]{2008Sci...321.1335S}
{Steinmetz}, T., {Wilken}, T., {Araujo-Hauck}, C., {Holzwarth}, R.,
  {H{\"a}nsch}, T.~W., {Pasquini}, L., {Manescau}, A., {D'Odorico}, S.,
  {Murphy}, M.~T., {Kentischer}, T., {Schmidt}, W., \& {Udem}, T., 2008.
\newblock {Laser Frequency Combs for Astronomical Observations}, {\it
  Science\/}, {\bf 321}(5894), 1335.

\bibitem[St{\"u}rmer et~al.(2017)St{\"u}rmer, Seifahrt, Schwab, \&
  Bean]{10.1117/1.JATIS.3.2.025003}
St{\"u}rmer, J., Seifahrt, A., Schwab, C., \& Bean, J.~L., 2017.
\newblock {Rubidium-traced white-light etalon calibrator for radial velocity
  measurements at the cm s-1 level}, {\it Journal of Astronomical Telescopes,
  Instruments, and Systems\/}, {\bf 3}(2), 025003.

\bibitem[{Tang} et~al.(2023){Tang}, {Hao}, {Ye}, {Zhai}, {Zhang}, \&
  {Xiao}]{2023AJ....165..156T}
{Tang}, L., {Hao}, Z., {Ye}, H., {Zhai}, Y., {Zhang}, K., \& {Xiao}, D., 2023.
\newblock {Drift Performance and Chromatic Thermal Response of a Temperature
  Stabilized Solid-etalon Calibrator}, {\it \aj\/}, {\bf 165}(4), 156.

\bibitem[{Terrien} et~al.(2021){Terrien}, {Ninan}, {Diddams}, {Mahadevan},
  {Halverson}, {Bender}, {Fredrick}, {Hearty}, {Jennings}, {Metcalf}, {Monson},
  {Roy}, {Schwab}, \& {Stef{\'a}nsson}]{2021AJ....161..252T}
{Terrien}, R.~C., {Ninan}, J.~P., {Diddams}, S.~A., {Mahadevan}, S.,
  {Halverson}, S., {Bender}, C., {Fredrick}, C., {Hearty}, F., {Jennings}, J.,
  {Metcalf}, A.~J., {Monson}, A., {Roy}, A., {Schwab}, C., \& {Stef{\'a}nsson},
  G., 2021.
\newblock {Broadband Stability of the Habitable Zone Planet Finder
  Fabry-P{\'e}rot Etalon Calibration System: Evidence for Chromatic Variation},
  {\it \aj\/}, {\bf 161}(6), 252.

\bibitem[{Wang} et~al.(2020){Wang}, {Wright}, {MacQueen}, {Cochran}, {Doss},
  {Gibson}, \& {Schmitt}]{2020PASP..132a4503W}
{Wang}, S.~X., {Wright}, J.~T., {MacQueen}, P., {Cochran}, W.~D., {Doss},
  D.~R., {Gibson}, C.~A., \& {Schmitt}, J.~R., 2020.
\newblock {Calibrating Iodine Cells for Precise Radial Velocities}, {\it
  \pasp\/}, {\bf 132}(1007), 014503.

\bibitem[Wildi et~al.(2010)Wildi, Pepe, Chazelas, Curto, \&
  Lovis]{10.1117/12.857951}
Wildi, F., Pepe, F., Chazelas, B., Curto, G.~L., \& Lovis, C., 2010.
\newblock {A Fabry-Perot calibrator of the HARPS radial velocity spectrograph:
  performance report}, in {\em Ground-based and Airborne Instrumentation for
  Astronomy III\/}, vol. 7735, p. 77354X, International Society for Optics and
  Photonics, SPIE.

\bibitem[{Wilken} et~al.(2010){Wilken}, {Lovis}, {Manescau}, {Steinmetz},
  {Pasquini}, {Lo Curto}, {H{\"a}nsch}, {Holzwarth}, \&
  {Udem}]{2010MNRAS.405L..16W}
{Wilken}, T., {Lovis}, C., {Manescau}, A., {Steinmetz}, T., {Pasquini}, L., {Lo
  Curto}, G., {H{\"a}nsch}, T.~W., {Holzwarth}, R., \& {Udem}, T., 2010.
\newblock {High-precision calibration of spectrographs}, {\it \mnras\/}, {\bf
  405}(1), L16--L20.

\bibitem[{Wilken} et~al.(2012){Wilken}, {Curto}, {Probst}, {Steinmetz},
  {Manescau}, {Pasquini}, {Gonz{\'a}lez Hern{\'a}ndez}, {Rebolo}, {H{\"a}nsch},
  {Udem}, \& {Holzwarth}]{2012Natur.485..611W}
{Wilken}, T., {Curto}, G.~L., {Probst}, R.~A., {Steinmetz}, T., {Manescau}, A.,
  {Pasquini}, L., {Gonz{\'a}lez Hern{\'a}ndez}, J.~I., {Rebolo}, R.,
  {H{\"a}nsch}, T.~W., {Udem}, T., \& {Holzwarth}, R., 2012.
\newblock {A spectrograph for exoplanet observations calibrated at the
  centimetre-per-second level}, {\it \nat\/}, {\bf 485}(7400), 611--614.

\bibitem[{Zechmeister, M.} et~al.(2018){Zechmeister, M.}, {Reiners, A.},
  {Amado, P. J.}, {Azzaro, M.}, {Bauer, F. F.}, {Béjar, V. J. S.}, {Caballero,
  J. A.}, {Guenther, E. W.}, {Hagen, H.-J.}, {Jeffers, S. V.}, {Kaminski, A.},
  {Kürster, M.}, {Launhardt, R.}, {Montes, D.}, {Morales, J. C.},
  {Quirrenbach, A.}, {Reffert, S.}, {Ribas, I.}, {Seifert, W.}, {Tal-Or, L.},
  \& {Wolthoff, V.}]{refId0}
{Zechmeister, M.}, {Reiners, A.}, {Amado, P. J.}, {Azzaro, M.}, {Bauer, F. F.},
  {Béjar, V. J. S.}, {Caballero, J. A.}, {Guenther, E. W.}, {Hagen, H.-J.},
  {Jeffers, S. V.}, {Kaminski, A.}, {Kürster, M.}, {Launhardt, R.}, {Montes,
  D.}, {Morales, J. C.}, {Quirrenbach, A.}, {Reffert, S.}, {Ribas, I.},
  {Seifert, W.}, {Tal-Or, L.}, \& {Wolthoff, V.}, 2018.
\newblock Spectrum radial velocity analyser (serval) - high-precision radial
  velocities and two alternative spectral indicators, {\it A\&A\/}, {\bf 609},
  A12.

\end{thebibliography}





\bsp	
\label{lastpage}
\end{document}